\documentclass[aps,prl,twocolumn]{revtex4}

\usepackage{amsmath,amssymb,epsfig,amsfonts}
\usepackage{graphicx}
\usepackage{times}

\begin{document}

\title{A {\it cryofuge\/} for cold-collision experiments with\\ slow polar molecules} 
\author{Xing Wu}
\email{xing.wu@yale.edu}
\altaffiliation[Current address: ]{Department of Physics, Yale University, New Haven, Connecticut 06520, USA.}
\author{Thomas Gantner}
\author{Manuel Koller}
\author{Martin Zeppenfeld}
\author{Sotir Chervenkov}
\author{Gerhard Rempe}

\affiliation{Max-Planck-Institut f\"ur Quantenoptik, Hans-Kopfermann-Str. 1, D-85748 Garching, Germany}

\begin{abstract}
Ultracold molecules represent a fascinating research frontier in physics and chemistry, but it has proven challenging to prepare dense samples at low velocities. Here we present a solution to this goal by a non-conventional approach dubbed cryofuge. It employs centrifugal force to bring cryogenically cooled molecules to kinetic energies below $1$K$\times k_B$ in the laboratory frame, with corresponding fluxes exceeding $10^{10}$/s at velocities below $20$\,m/s. By attaining densities higher than $10^9$/cm$^3$ and interaction times longer than $25$\,ms in samples of fluoromethane as well as deuterated ammonia, we observe cold dipolar collisions between molecules and determine their collision cross sections.
\end{abstract}



\maketitle

Ultracold atoms have become established workhorses in several fields of physics, and molecule slowing and cooling methods are now coming to the fore. Compared to atoms, molecules possess a variety of unique properties like permanent electric dipole moments. These offer, on the one hand, a convenient handle to experimentally manipulate collision pathways, an intriguing vista for the emerging field of cold and ultracold chemistry~\cite{Herschbach2009,Ospelkaus2010,Quéméner2012}. On the other hand, the long-range and anisotropic dipole coupling mediates interactions over micrometer distances~\cite{Yan2013}. This renders cold polar molecules particularly suitable for applications in quantum simulation~\cite{Baranov2012} and computing~\cite{DeMille2002} such as investigation of strongly correlated and dipole blockaded systems.

The generic roadmap towards the envisioned experiments is to prepare molecular samples that are cold, dense, and slow. These three conditions respectively ensure that quantum phenomena can be distinguished from thermal physics, interactions between molecules are frequent, and observation times are long. Moreover, these criteria constitute a prerequisite for sympathetic~\cite{Lara2006} and evaporative cooling of trapped molecules~\cite{Reens2017} en route towards quantum degeneracy. So far, only a special class of alkali dimers associated from atoms can meet the three conditions simultaneously~\cite{Ni2008,Park2015}. However, notwithstanding the remarkable progress in the past decade, experiments with naturally occurring molecules have not yet reached this regime. Existing methods to manipulate molecules yield cold and dense but fast samples~\cite{Tizniti2014,Henson2012,Jankunas2015}, or cold and slow but dilute samples~\cite{Bethlem1999,Cheng2016,Norrgard2016,Prehn2016,Truppe2017,Liu2017}. Finding a general approach to produce cold, dense, and slow molecules therefore constitutes an outstanding challenge.

Here we present a non-conventional solution dubbed cryofuge that delivers a guided beam of cold and slow molecules. Cryofuge denotes successive cryogenic buffer-gas cooling~\cite{Weinstein1998,vanBuuren2009,Hutzler2012} and centrifuge deceleration~\cite{Chervenkov2014}. As neither step involves the specific internal structure of the molecules, this renders the method generic. We demonstrate the capabilities of the cryofuge by showing internal-state cooling and deceleration of fluoromethane (CH$_3$F). We illustrate the generality of the method by extension to several other compounds, including ammonia (ND$_3$), methanol (CH$_3$OH), (trifluoromethyl)acetylene (CF$_3$CCH), and isopropanol (C$_3$H$_7$OH). The high intensities achieved at laboratory velocities below 20m/s allow for molecule interaction times exceeding $25\,$ms, and thus enable the observation of cold dipolar collisions with a measured collision rate of $\sim10\,$Hz, which we demonstrate with CH$_3$F and ND$_3$. Such a rate is several hundred times larger than the loss rate of molecules from state-of-the-art electric traps~\cite{Meerakker2012,Englert2011,Prehn2016}, and provides perfect starting conditions for further collision and possibly evaporative-cooling~\cite{Reens2017} experiments. The measured large scattering lengths, characteristic of dipolar interactions, also offer access to strongly correlated systems~\cite{Baranov2012} with seconds or even minutes-long lifetimes in a room-temperature environment.

\begin{figure}
	\centering
		\includegraphics[width=0.5\textwidth]{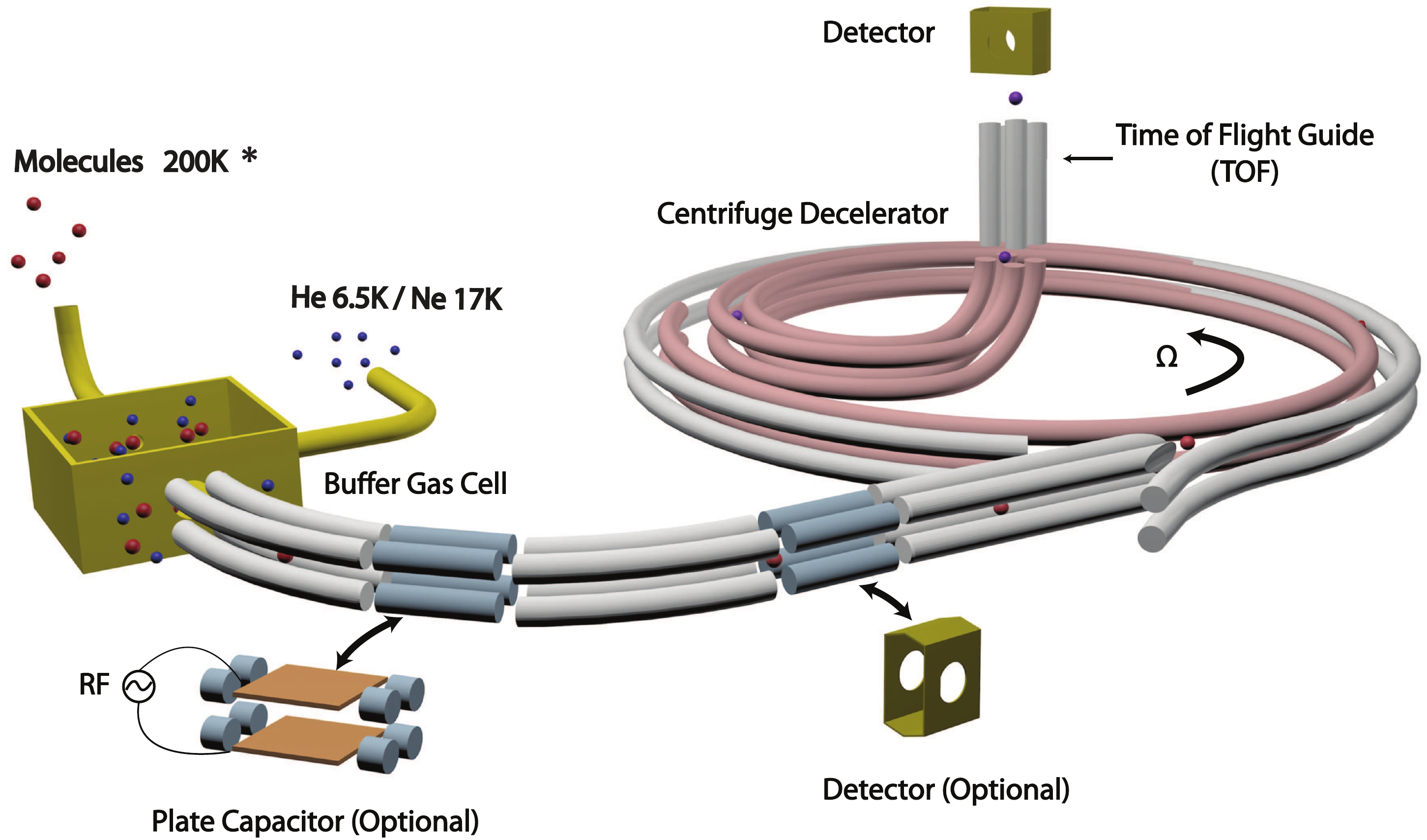}
	\caption{
	\textbf{Schematic illustration of the cryofuge (not to scale).} A cryogenic buffer-gas cell and a centrifuge decelerator are connected by a quadrupole electrostatic guide. The buffer-gas cell operates with either He at $6.5$\,K, or Ne at $17$\,K. The centrifuge decelerator consists of a pair of circular static electrodes (radius=$20$\,cm) around the periphery (silver) and a spiral-shaped rotating quadrupole guide (red). The inner pair of rotating electrodes is extended to form a storage ring with the static electrodes around the periphery, enabling the deceleration of continuous beams. The optional components, a parallel-plate capacitor and a detector, can replace the straight guiding segments (blue), when characterizing the internal-state distribution of the buffer-gas-cooled molecules by resonant radio-frequency depletion spectroscopy. $^\ast$ Methanol was introduced at $400$\,K.
	}
	\label{fig:setup}
\end{figure}
The operating principle of the cryofuge is illustrated in Fig.\,1. Warm molecules first thermalize with a helium or neon buffer-gas in the cryogenic buffer-gas cell. Cell operation is possible in two density regimes, a lower-pressure boosted regime and a higher-pressure supersonic regime~\cite{Hutzler2012,Wu2016}, resulting in either higher total flux or higher state-purity, respectively. Molecules are extracted from the cryogenic environment by a bent electrostatic quadrupole guide (radius of curvature $20$\,cm), and transferred by a straight guide to the centrifuge decelerator. Here, the molecules enter a rotating guide wherein the centrifugal potential slows them down almost to a standstill. A final straight guide brings the molecules to a quadrupole mass spectrometer (QMS) for detection.

Inserting an optional parallel-plate capacitor with radio-frequency strip electrodes and a removable quadrupole mass spectrometer (QMS) into the guide before the centrifuge enables probing of rotational states by resonant depletion spectroscopy, as described in~\cite{Wu2016}. Moreover, the straight guide between the centrifuge output and the QMS detector can be toggled on and off to determine longitudinal velocities ($v_z$) via time-of-flight (TOF) measurements. Varying the guiding voltage, hence the trap depth, on the straight guide before the centrifuge can control molecular densities for cold collision measurements after deceleration.

\begin{figure}
	\centering
		\includegraphics[width=0.42\textwidth]{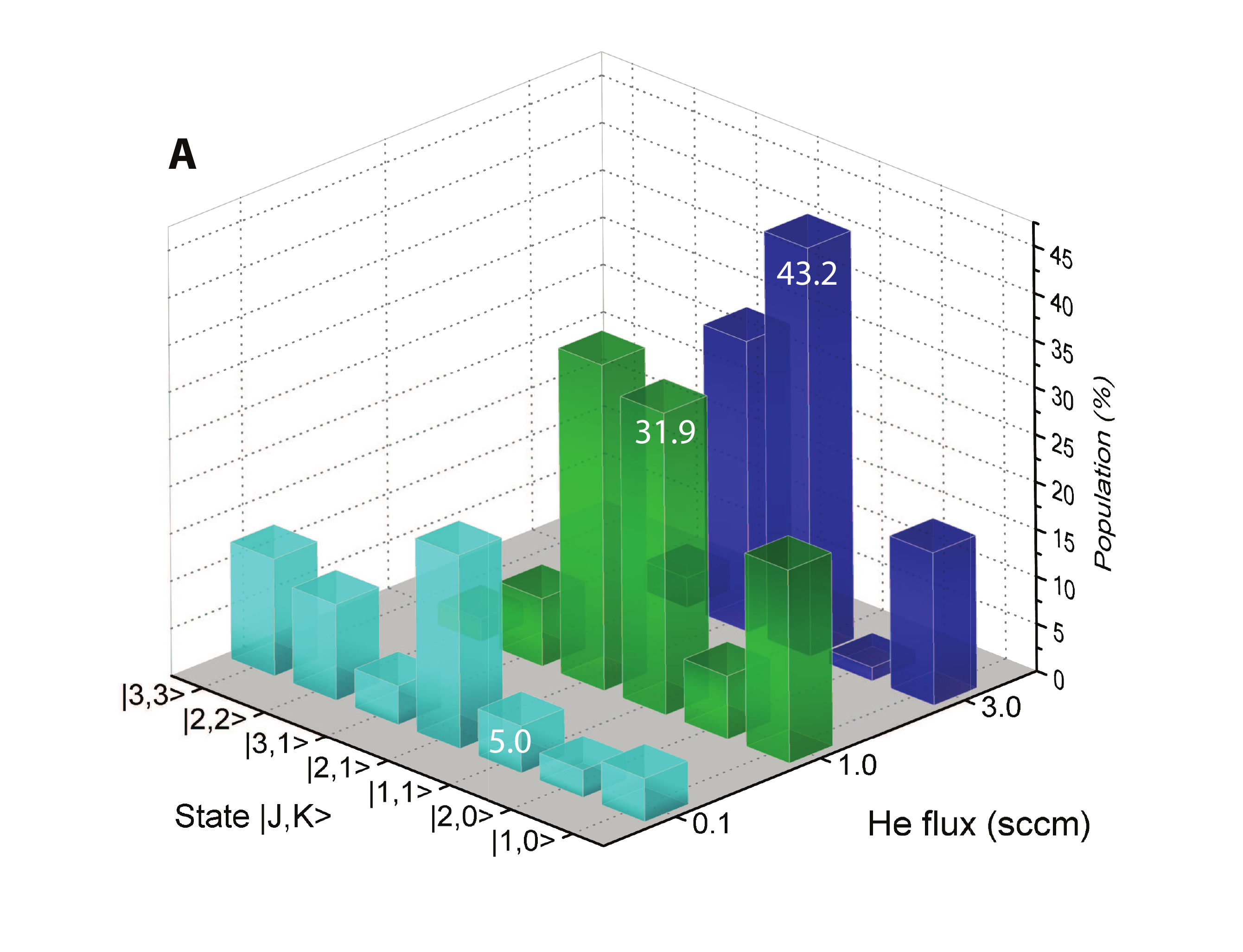}
		\includegraphics[width=0.42\textwidth]{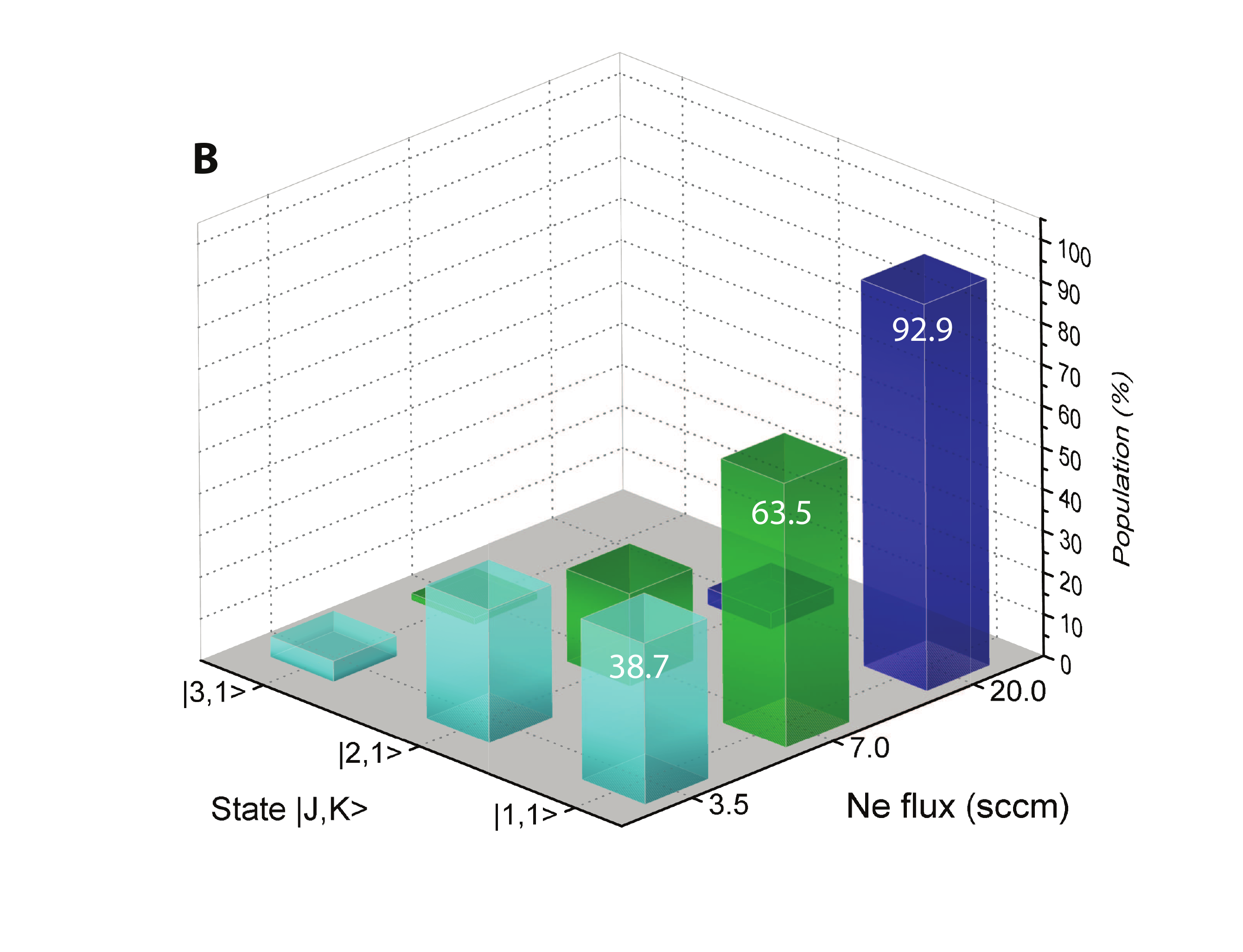}

	\caption{
	\textbf{Measured rotational-state mapping of CH$_3$F in an electrostatic guide, as a function of the buffer-gas flux.} (\textbf{A}) Buffer-gas cooling in the boosted regime (with a moderate He inflow at $6.5$\,K). Unit of gas flow: $1$ sccm $=4.1\times10^{17}$\,s$^{-1}$. The left axis labels the rotational states $|$J,K$\rangle$ in ascending order in energy. The 7 lowest states with substantial guided population are shown. The vertical scale shows the relative population of each probed state in the guide. (\textbf{B}) Buffer-gas cooling in the supersonic regime (with a much higher Ne inflow at $17$\,K). The errorbars of the measured populations in (\textbf{A}) and (\textbf{B}) are all below $0.6$\%, and are not visible in the plots.
	}
	\label{fig:statesdistr}
\end{figure}

We first studied rotational-state cooling of CH$_3$F, tracking the variation of the internal state distribution as a function of the buffer-gas density for the boosted (Fig.\,2A) and the supersonic (Fig.\,2B) regimes. In both cases a higher buffer-gas density results in a higher state purity, with an increasing population in the $|$J,K$\rangle=|$1,1$\rangle$ ground level of the para nuclear spin states. We achieved a maximum population of $(43.2\pm0.3)$\,\% and $(92.9\pm0.5)$\,\% in the $|$1,1$\rangle$ state of CH$_3$F in the boosted and supersonic regime, respectively.

\begin{figure}[ht]
	\centering
		\includegraphics[width=0.50\textwidth]{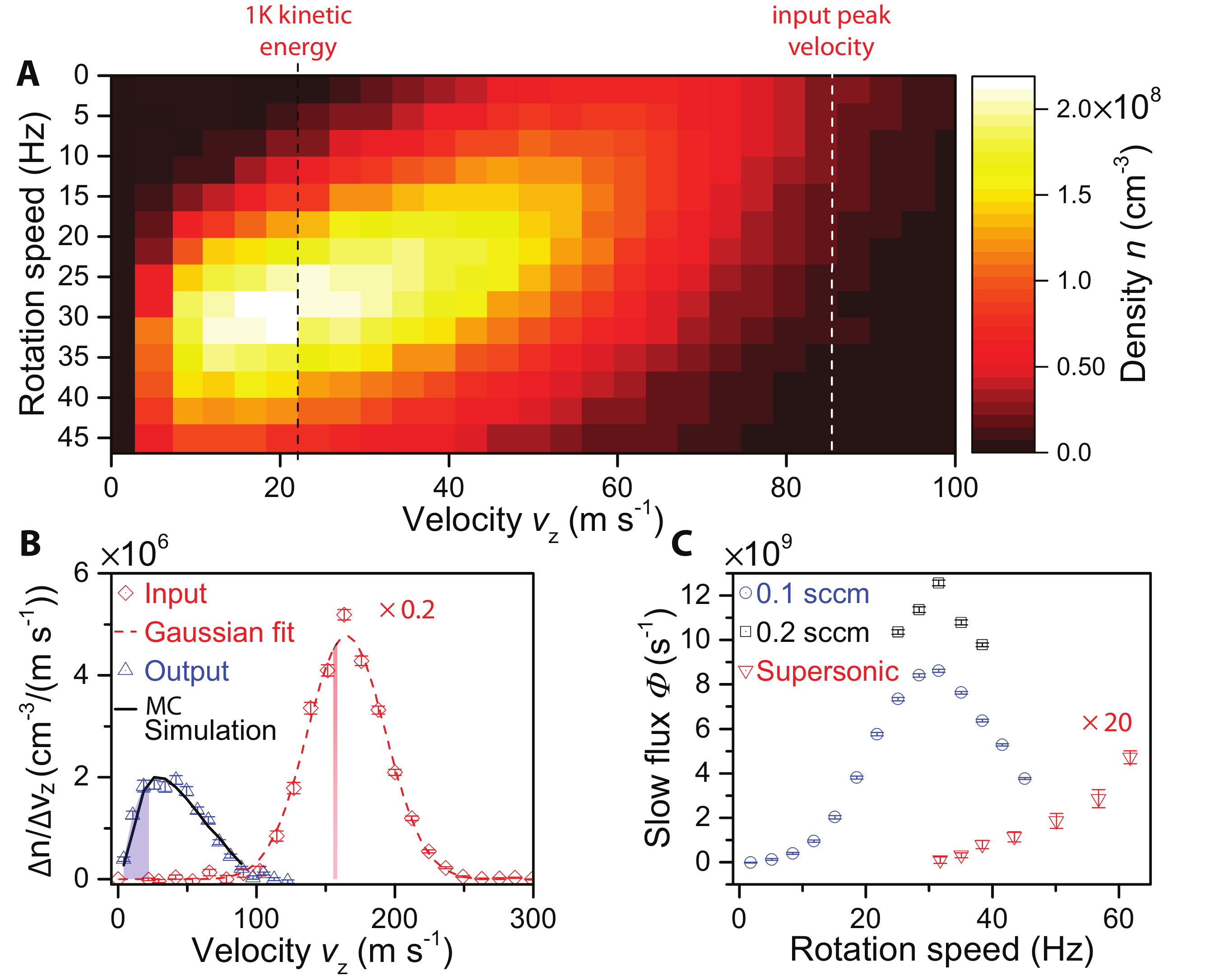}
	\caption{
	\textbf{Results from the centrifuge deceleration of CH$_3$F.} (\textbf{A}) 2D map of the output longitudinal velocity ($v_z$) distribution of molecules from a boosted cell ($6.5$\,K buffer-gas cell, $0.5$\,sccm He inflow, $0.1$\,sccm molecule inflow). The vertical axis shows the centrifuge rotation speed. The color scale indicates the molecular density in each velocity interval. The white dashed line at $86$\,m/s marks the peak velocity of the input molecular beam. The black dashed line corresponds to $1$K kinetic energy for CH$_3$F. (\textbf{B}) Comparison between the input (rescaled by a factor of 0.2) and output $v_z$-distributions from a supersonic cell ($17$\,K buffer-gas cell, $7$\,sccm Ne inflow, $0.1$\,sccm molecule inflow) at $62$\,Hz centrifuge rotation. A Gaussian curve is fitted (least-$\chi^2$ method) to the input with its center, width, and amplitude as free parameters. The vertical axis shows density per velocity interval. (\textbf{C}) Output slow flux ($1$K$\times k_B$ kinetic energy) versus centrifuge rotation speed. The blue and black plots depict results from the boosted regime ($6.5$\,K cell, $0.5$\,sccm He inflow) at different molecule input ($0.1$ and $0.2$\,sccm, respectively). The red plot (rescaled by a factor of 20) depicts the supersonic results ($17$\,K buffer-gas cell, $7$\,sccm Ne inflow, $0.1$\,sccm molecule inflow). The errorbars in (\textbf{B}) and (\textbf{C}) represent $1\sigma$ statistical error, derived from signal shot noise averaged over typically $15$\, minutes measurement time.
	}
	\label{fig:velocitydistr}
\end{figure}
An inherent problem of buffer-gas cooling in both these regimes is that slow molecules are missing in the extracted samples due to collisions with the buffer-gas in the vicinity of the nozzle. This is resolved with the centrifuge. Figure\,3A illustrates the deceleration of CH$_3$F for cell operation in the boosted regime. The two-dimensional map shows the variation of the output $v_z$-distributions (horizontal axis) as a function of the centrifuge rotation speed (vertical axis). Two effects are apparent. First, the peak of the $v_z$-distribution shifts towards lower velocities with increasing rotation speed, as expected for deceleration. Second, the molecular density increases for rotation speeds below $30$\,Hz, and then decreases. The increase comes from deceleration, with more molecules surviving the small bend (radius 5cm) at the exit of the centrifuge. The decrease of density towards higher rotation speeds is due to over-deceleration, with the molecules no longer having sufficient kinetic energy to climb the centrifugal potential and reach the output of the centrifuge~\cite{Chervenkov2014}. 

The deceleration of a beam in the supersonic regime is shown in Fig.\,3B, where the measured $v_z$-distributions at both the input and the output of the centrifuge are compared. The fit to the input distribution indicates a peak velocity of $165$\,m/s and a velocity spread corresponding to $3.3\pm0.1$\,K in the co-moving frame. By rotating the centrifuge at $62$\,Hz, the $v_z$-interval near the peak of the input distribution (red stripe) is shifted to below $1$K$\times k_B$ kinetic energy (blue area) in the lab. As the vertical scale $\Delta n/\Delta v_z$ is proportional to the phase space density, the ratio of the two heights gives the output efficiency of the centrifuge deceleration, which is about $8$\% for the supersonic input. For a boosted input, the efficiency improves to typically $20$\% due to the lower input velocities and hence better electrostatic guiding in the centrifuge. The decrease of signal towards zero velocities is due to the divergence of the very slow molecules in the gap between the centrifuge output and the TOF guide, as confirmed by Monte-Carlo trajectory simulations. 

A key figure of merit is the flux of slow ($<1$K$\times k_B$) molecules obtained. Such molecules are required for trap loading as well as for cold-collision studies and cold-chemistry experiments with long interaction times. Figure\,3C shows the flux of slow molecules obtained in both the boosted (with $0.1$ and $0.2$\,sccm molecule input) and the supersonic regime. At $0.2$\,sccm molecule input in the boosted regime, we have obtained a slow-molecule flux of $(1.2\pm^{1.2}_{0.6})\times10^{10}$\,s$^{-1}$ with a peak density of $(1.0\pm^{1.0}_{0.5})\times10^9$\,cm$^{-3}$ at $30$\,Hz rotation. The factor-of-two error results from systematics in the QMS sensitivity calibration. Compared to the first demonstration of centrifuge deceleration~\cite{Chervenkov2014}, the current results represent an enhancement by at least $1$ order of magnitude in the obtained slow-molecule flux and density alone, and about $3$ orders of magnitude if population in a single  quantum state is taken into account. The phase-space density is estimated to be $\sim5\times10^{-14}$, comparable to the value achieved for a molecular magneto-optical trap~\cite{Norrgard2016}, but now already for temperatures around $1$\,K instead of $<1$\,mK. In comparison, the flux of slow molecules obtained in the supersonic regime is about $40$ times smaller, mainly due to the limited buffer-gas cell extraction and the lower guiding efficiency.

The generic principle underlying the cryofuge enables its application to a wide range of compounds. For ND$_3$ and CF$_3$CCH which are symmetric-tops like CH$_3$F, fluxes of $\sim1\times10^{10}$\,s$^{-1}$ and densities of $\sim1\times10^{9}$\,cm$^{-3}$ have been obtained below $1$K$\times k_B$ kinetic energy. We have also applied the cryofuge to methanol and isopropanol. For methanol, a flux of $\sim3\times10^8$\,s$^{-1}$ at a peak velocity of $40$\,m~s$^{-1}$ and $\sim1\times10^7$\,s$^{-1}$ below $1$K$\times k_B$ was produced (Fig.\,S2). The comparably lower flux results from both the weaker Stark shifts~\cite{Jansen2013} and the much lower vapor pressure of methanol. We have achieved a similar flux of $\sim1\times10^8$\,s$^{-1}$ for isopropanol, but maintaining stable operation requires additional effort, as isopropanol is prone to freezing due to its even lower vapor pressure. 

The high density of the decelerated molecules brings us into the cold-collision regime, which we investigated for CH$_3$F and ND$_3$. We observed increasing losses for decreasing velocities in the molecular beams (Fig.\,4), and attribute this to dipolar collisions mainly in the TOF-guide. 

\begin{figure*}[ht]
	\centering
		\includegraphics[width=0.65\textwidth]{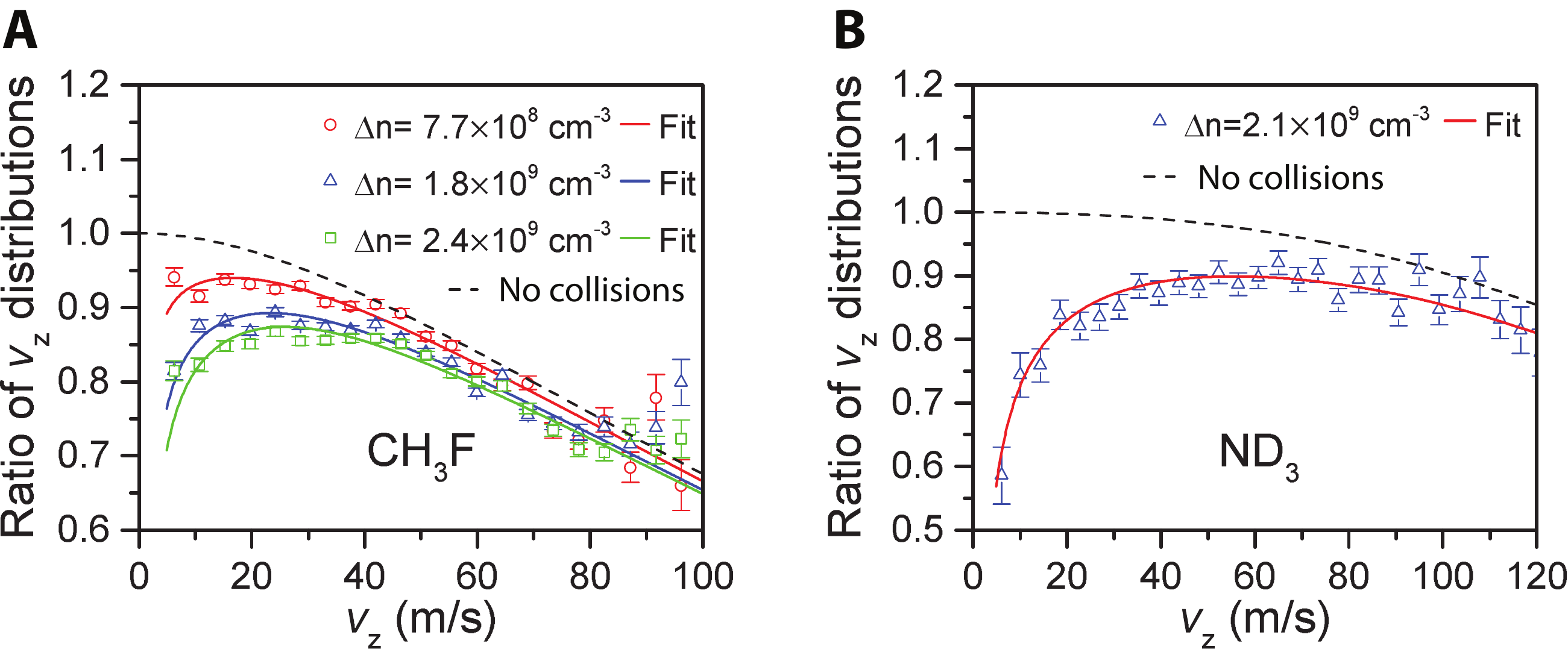}
	\caption{
	\textbf{Depletion from molecule-molecule collisions in the centrifuge-decelerated beam.} (\textbf{A}) Ratio of $v_z$-distributions for high and low CH$_3$F densities, taken always at $10$\,kV and $1$\,kV, respectively, on the straight guide before the centrifuge. The three $\Delta n$ values correspond to $0.06$, $0.2$, and $0.3$\,sccm CH$_3$F input flux (in descending order), at $6.5$\,K cell temperature, $0.5$\,sccm He inflow, and $33.5$\,Hz centrifuge rotation speed. The corresponding $1\sigma$ statistical errorbars are derived from signal shot noise averaged over $12$, $6$, and $4$ hours, respectively.  (\textbf{B}) Ratio of $v_z$-distributions for high and low ND$_3$ densities, at $8.5$\,K cell temperature, $0.5$\,sccm He inflow, $0.2$\,sccm ND$_3$ inflow, and $38.5$\,Hz centrifuge rotation. The data are averaged over $8$ hours measurement. For both measurements, the $|$1,1$\rangle$ state has the largest population, a measured $43$\% for CH$_3$F and an estimated $80$\% for ND$_3$.The black dashed lines in both figures show the ratio of simulated $v_z$-distributions for the collision-free case.
	}
	\label{fig:collision}
\end{figure*}
To quantify the collisional loss, we examined its dependence on the molecular density $n$ and the longitudinal velocity $v_z$ of the guided beam, as well as on the averaged dipole moment $\left\langle d\right\rangle$ of the molecule. The $v_z$-dependence occurs because the collision probability is proportional to the passage time through the guide, hence $\propto1/v_z$.
Figure\,4 shows the normalized ratios of $v_z$-distributions taken at high and low molecule densities corresponding to a density difference $\Delta n$, set by varying the voltage on the straight guide before the centrifuge. In addition, for a fixed voltage difference, $\Delta n$ can be tuned by regulating the molecule inflow to the cryosource. For guided beams, the collisional loss modifies the original $v_z$-distribution $P_0$ into $P=P_0exp(-k_{loss}nL /v_z)$, where $k_{loss}=\sigma_{loss}v_{rel}$ is the loss rate coefficient with $\sigma_{loss}$ being the total collision cross section for all loss channels and $v_{rel}$ being the relative velocity between colliding partners, and $L =L_{TOF}+L_{eff}$ is the guide length. $L$ includes the contribution from the TOF-guide $L_{TOF}=46$\,cm, and additionally an effective guide length inside the centrifuge $L_{eff}$. The latter takes into account the variation of $v_z$ and $n$ during deceleration and depends on the exit velocity from the centrifuge (Fig.\,S7\,C and D). Thus the ratio of two $v_z$-distributions $P_{High}/P_{Low}$ for a $\Delta n=n_{High}-n_{Low}$ can be fitted with an $exp(-\alpha k_{th}\Delta nL/v_z)$-model, where $k_{th}$ is the theoretically predicted loss rate coefficient, and $\alpha$ is the only fit parameter which accounts for the deviation between theory and the experimentally measured loss. Moreover, the fit includes a collision-independent background, which is well-understood and results from the electrostatic filtering in the bent guide (Fig.\,S3).

The theoretical rate coefficient includes losses from both elastic and inelastic contributions, $k_{th}=k^{el}_{loss}+k^{in}$, which are estimated with the semiclassical eikonal approximation~\cite{SakuraiQM} and the Langevin capture model~\cite{Bell2009}, respectively. Not all elastic collisions contribute to losses, but only those that increase the transverse energy to a value exceeding the maximum guiding potential. With this picture in mind, we compute the probability for a molecule to be lost for a given collision event with scattering angle $\theta$, relative longitudinal velocity $v_{z}^{rel}$, and given relative transverse velocity, and radial position where the collision occurs. We then integrate this loss probability over the transverse velocity and spatial distribution to obtain $P_{loss}(v_{z}^{rel},\theta)$ (Fig.\,S5\,A). We also calculate the differential cross-section $\frac{d\sigma}{d\Omega}(v_{z}^{rel},\theta)$ (Fig.\,S5\,C) for the dipole interaction. The elastic loss cross section is then $\sigma^{el}_{loss}(v_{z}^{rel})=2\pi\int d\theta sin\theta P_{loss}\frac{d\sigma}{d\Omega}$. For the inelastic loss, it has been suggested that the presence of an external electric field could induce pronounced orientation-changing collisions between polar molecules~\cite{Bohn2001,Parazzoli2011}. As an estimate, we assume the Langevin model accounts for all inelastic losses and use it to compute $k^{in}$. A comparison of these $v_{z}^{rel}$-dependent rate coefficients is presented in the supplement.

With the averaged dipole $\left\langle d_{CH_3F}\right\rangle=0.56$\,D and $\left\langle d_{ND_3}\right\rangle=0.77$\,D in our decelerated beams, we obtain the theoretical values $k_{th}^{CH_3F}=7.7\times10^{-10}$\,cm$^3$s$^{-1}$ and $k_{th}^{ND_3}=1.3\times10^{-9}$\,cm$^3$s$^{-1}$ at $0.8$K$\times k_B$ and $1.1$K$\times k_B$ collision energy, respectively. The fit to the data reveals $\alpha^{CH_3F}=1.4\pm0.1$ and $\alpha^{ND_3}=1.6\pm0.3$, meaning that the measured loss rate coefficients are between $40$\% and $60$\% larger than theory, respectively. This deviation can be attributed to our underestimation of molecular density in the centrifuge, and the imprecision of the Langevin model. We note that although the total elastic cross sections are large, calculated to be $\sigma^{el}_{CH_3F}=2.0\times10^{-12}$\,cm$^2$ and $\sigma^{el}_{ND_3}=2.5\times10^{-12}$\,cm$^2$, the loss ratio is small with $\sigma^{el}_{loss}/\sigma^{el}_{total}\approx6$\% for both cases. Only $1$ out of $17$ elastic collisions leads to loss.

For dipolar collisions, the semiclassical calculations~\cite{Bohn2009,Cavagnero2009} predict a rate coefficient $k\propto\left\langle d\right\rangle^2$.
Hence we expect $k^{ND_3}/k^{CH_3F}=\left\langle d_{ND_3}\right\rangle^2/\left\langle d_{CH_3F}\right\rangle^2$ to be approximately $1.9$.
This agrees nicely with the experimental value $k_{loss}^{ND_3}/k_{loss}^{CH_3F}=1.9\pm0.4$. The systematic error from the density calibration cancels out in this ratio for reasons explained in the supplement.

Finally, to show that we indeed explore the onset of the huge and long-range dipolar interaction between cold polar molecules, we compare its strength $V_{dd}(r)=-\left\langle d\right\rangle^2/4\pi\epsilon_0r^3$ to that of the van der Waals interaction $V_{vdW}(r)=-C_6/r^6$ where $C_6$ is the dispersion coefficient and $r$ is the inter-molecular distance. The $C_{6}$ values for CH$_3$F-CH$_3$F and ND$_3$-ND$_3$ collisions estimated from the London formula~\cite{Atkins2005} are $100$\,a.u. and $55$\,a.u. ($1$\,a.u.$=1E_ha_0^6$, where $E_h=4.36\times10^{-18}$\,J and $a_0=0.529$\,\AA), respectively. At the interaction distance $r_0=\sqrt{\sigma_{loss}/\pi}\sim4$\,nm for $\sigma_{loss}\sim0.5\times10^{-12}$\,cm$^2$ (Table\,S4), $V_{dd}(r_0)/V_{vdW}(r_0)\sim200$, and thus the dipole-dipole interaction dominates by far.

The measured molecular density and the elastic collision rate give a collision frequency of $\sim10$\,Hz, which would imply a thermalization rate of $\sim1$\,Hz~\cite{DavisEvaporative}. Thus, once combined with the previously developed and readily compatible electrostatic trap  featuring minute-long trapping times~\cite{Englert2011,Prehn2016}, it becomes possible to measure thermalization rates for different internal states and collision energies as well as various molecular species. Furthermore, implementing optoelectrical Sisyphus cooling~\cite{Zeppenfeld2009} to reach a lower temperature in such a trap might allow us to explore the scattering resonance at the threshold limit~\cite{Bohn2009} and the giant dipolar collision cross sections in or close to the quantum regime, and possibly even implement evaporative cooling of polyatomic molecules.

In addition to enabling the observation of cold dipolar collisions, the broad versatility of the cryofuge could drastically extend the scope of current cold-molecule research. For instance, the cold and slow beam of methanol we produced could be ideal for measuring variations in the proton-to-electron mass ratio~\cite{Bagdonaite2013}. Moreover, the cryofuge could serve as a perfect source for ongoing experiments with laser-cooled diatomic molecules~\cite{Norrgard2016,Truppe2017}, as many of the species considered there possess rotational states with Stark shifts sufficiently large to apply our cryofuge technique.

We thank Tobias Urban, Florian Furchtsam, Johannes Siegl, and Ferdinand Jarisch for technical support.

\bibliographystyle{unsrt}

\begin{thebibliography}{10}

\bibitem{Herschbach2009}
D.~Herschbach, {\it Faraday Discuss.\/} {\bf 142}, 9 (2009).

\bibitem{Ospelkaus2010}
S.~Ospelkaus, {\it et~al.\/}, {\it Science\/} {\bf 327}, 853 (2010).

\bibitem{Quéméner2012}
G.~Qu\'em\'ener, P.~S. Julienne, {\it Chem. Rev.\/} {\bf 112}, 4949 (2012).

\bibitem{Yan2013}
B.~Yan, {\it et~al.\/}, {\it Nature\/} {\bf 501}, 521 (2013).

\bibitem{Baranov2012}
M.~A. Baranov, M.~Dalmonte, G.~Pupillo, P.~Zoller, {\it Chem. Rev.\/} {\bf
  112}, 5012 (2012).

\bibitem{DeMille2002}
D.~DeMille, {\it Phys. Rev. Lett.\/} {\bf 88}, 067901 (2002).

\bibitem{Lara2006}
M.~Lara, J.~L. Bohn, D.~Potter, P.~Sold\'an, J.~M. Hutson, {\it Phys. Rev.
  Lett.\/} {\bf 97}, 183201 (2006).

\bibitem{Reens2017}
D.~L. Reens, H.~Wu, T.~Langen, J.~Ye, {\it arXiv:1706.02806v1\/}  (2017).

\bibitem{Ni2008}
K.-K. Ni, {\it et~al.\/}, {\it Science\/} {\bf 322}, 231 (2008).

\bibitem{Park2015}
J.~W. Park, S.~A. Will, M.~W. Zwierlein, {\it Phys. Rev. Lett.\/} {\bf 114},
  205302 (2015).

\bibitem{Tizniti2014}
M.~Tizniti, {\it et~al.\/}, {\it Nature Chemistry\/} {\bf 6}, 141 (2014).

\bibitem{Henson2012}
A.~B. Henson, S.~Gersten, Y.~Shagam, J.~Narevicius, E.~Narevicius, {\it
  Science\/} {\bf 338}, 234 (2012).

\bibitem{Jankunas2015}
J.~Jankunas, K.~Jachymski, M.~Hapka, A.~Osterwalder, {\it J. Chem. Phys.\/}
  {\bf 142}, 164305 (2015).

\bibitem{Bethlem1999}
H.~L. Bethlem, G.~Berden, G.~Meijer, {\it Phys. Rev. Lett.\/} {\bf 83}, 1558
  (1999).

\bibitem{Cheng2016}
C.~Cheng, {\it et~al.\/}, {\it Phys. Rev. Lett.\/} {\bf 117}, 253201 (2016).

\bibitem{Norrgard2016}
E.~B. Norrgard, D.~J. McCarron, M.~H. Steinecker, M.~R. Tarbutt, D.~DeMille,
  {\it Phys. Rev. Lett.\/} {\bf 116}, 063004 (2016).

\bibitem{Prehn2016}
A.~Prehn, M.~Ibr\"ugger, R.~Gl\"ockner, G.~Rempe, M.~Zeppenfeld, {\it Phys.
  Rev. Lett.\/} {\bf 116}, 063005 (2016).

\bibitem{Truppe2017}
S.~Truppe, {\it et~al.\/}, {\it Nature Physics\/} doi:10.1038/nphys4241  (2017).

\bibitem{Liu2017}
Y.~Liu, {\it et~al.\/}, {\it Phys. Rev. Lett.\/} {\bf 118}, 093201 (2017).

\bibitem{Weinstein1998}
J.~Weinstein, R.~DeCarvalho, T.~Guillet, B.~Friedrich, J.~Doyle, {\it Nature\/}
  {\bf 395}, 148 (1998).

\bibitem{vanBuuren2009}
L.~D. van Buuren, {\it et~al.\/}, {\it Phys. Rev. Lett.\/} {\bf 102}, 033001
  (2009).

\bibitem{Hutzler2012}
N.~R. Hutzler, H.-I. Lu, J.~M. Doyle, {\it Chem. Rev.\/} {\bf 112}, 4803
  (2012).

\bibitem{Chervenkov2014}
S.~Chervenkov, {\it et~al.\/}, {\it Phys. Rev. Lett.\/} {\bf 112}, 013001
  (2014).

\bibitem{Meerakker2012}
S.~Y.~T. van~de Meerakker, H.~L. Bethlem, N.~Vanhaecke, G.~Meijer, {\it Chem.
  Rev.\/} {\bf 112}, 4828 (2012).

\bibitem{Englert2011}
B.~G.~U. Englert, {\it et~al.\/}, {\it Phys. Rev. Lett.\/} {\bf 107}, 263003
  (2011).

\bibitem{Wu2016}
X.~Wu, T.~Gantner, M.~Zeppenfeld, S.~Chervenkov, G.~Rempe, {\it ChemPhysChem\/}
  {\bf 17}, 3631 (2016).

\bibitem{Jansen2013}
P.~Jansen, {\it et~al.\/}, {\it Mol. Phys.\/} {\bf 111}, 1923 (2013).

\bibitem{SakuraiQM}
J.~Sakurai, J.~J. Napolitano, {\it Modern quantum mechanics\/} (PEARSON
  Education Limited, 2014), second edn.

\bibitem{Bell2009}
M.~T. Bell, T.~P. Softley, {\it Mol. Phys.\/} {\bf 107}, 99 (2009).

\bibitem{Bohn2001}
J.~L. Bohn, {\it Phys. Rev. A\/} {\bf 63}, 052714 (2001).

\bibitem{Parazzoli2011}
L.~P. Parazzoli, N.~J. Fitch, P.~S. \ifmmode~\dot{Z}\else \.{Z}\fi{}uchowski,
  J.~M. Hutson, H.~J. Lewandowski, {\it Phys. Rev. Lett.\/} {\bf 106}, 193201
  (2011).

\bibitem{Bohn2009}
J.~L. Bohn, M.~Cavagnero, C.~Ticknor, {\it New J. Phys.\/} {\bf 11}, 055039
  (2009).

\bibitem{Cavagnero2009}
M.~Cavagnero, C.~Newell, {\it New J. Phys.\/} {\bf 11}, 055040 (2009).

\bibitem{Atkins2005}
P.~Atkins, R.~Friedman, {\it Molecular Quantum Mechanics\/} (Oxford University
  Press, Oxford, 2005), fourth edn.

\bibitem{DavisEvaporative}
K.~B. Davis, M.-O. Mewes, M.~A. Joffe, M.~R. Andrews, W.~Ketterle, {\it Phys.
  Rev. Lett.\/} {\bf 74}, 5202 (1995).

\bibitem{Zeppenfeld2009}
M.~Zeppenfeld, M.~Motsch, P.~W.~H. Pinkse, G.~Rempe, {\it Phys. Rev. A\/} {\bf
  80}, 041401 (2009).
	

\bibitem{Bagdonaite2013}
J.~Bagdonaite, {\it et~al.\/}, {\it Science\/} {\bf 339}, 46 (2013).


\bibitem{Sommer2009}
C.~Sommer, {\it et~al.\/}, {\it Faraday Discuss.\/} {\bf 142}, 203 (2009).

\bibitem{MotschBoosting}
M.~Motsch, {\it et~al.\/}, {\it New J. Phys.\/} {\bf 11}, 055030 (2009).

\bibitem{Beijerinck1974}
H.~C.~W. Beijerinck, R.~G. J.~M. Moonen, N.~F. Verster, {\it J. Phys. E Sci.
  Instrum.\/} {\bf 7}, 31 (1974).

\bibitem{BRAUN1993}
R.~Braun, P.~Hess, {\it Int. J. Mass Spectrom. Ion Processes\/} {\bf 125}, 229
  (1993).

\bibitem{Sommer2010}
C.~Sommer, {\it et~al.\/}, {\it Phys. Rev. A\/} {\bf 82}, 013410 (2010).

\bibitem{Hwang1996}
W.~Hwang, Y.-K. Kim, M.~E. Rudd, {\it J. Chem. Phys.\/} {\bf 104}, 2956 (1996).

\bibitem{Rao1992}
M.~V. V.~S. Rao, S.~K. Srivastava, {\it J. Phys. B: At. Mol. Opt. Phys.\/} {\bf
  25}, 2175 (1992).

\bibitem{Torres2001}
I.~Torres, R.~Marti´nez, M.~N.~S. Rayo, F.~Castaño, {\it J. Chem. Phys.\/} {\bf
  115}, 4041 (2001).

\bibitem{Srivastava1996}
S.~K. Srivastava, E.~Krishnakumar, A.~F. Fucaloro, T.~van Note, {\it J.
  Geophys. Res. Planets\/} {\bf 101}, 26155 (1996).

\bibitem{nistdatabase}
Nist computational chemistry comparison and benchmark database,
  http://cccbdb.nist.gov/ (2016). NIST Standard Reference Database Number 101.

\bibitem{Lampe1957}
F.~W. Lampe, J.~L. Franklin, F.~H. Field, {\it J. Am. Chem. Soc.\/} {\bf 79},
  6129 (1957).


\end{thebibliography}

\end{document}


\title{{\Large Supplementary Material for: }\\\vspace{0.4cm}A cryofuge for cold-collision experiments with slow polar molecules}
\author{Xing Wu}
\email{xing.wu@yale.edu}
\altaffiliation[Current address: ]{Department of Physics, Yale University, New Haven, Connecticut 06520, USA.}
\author{Thomas Gantner}
\author{Manuel Koller}
\author{Martin Zeppenfeld}
\author{Sotir Chervenkov}
\author{Gerhard Rempe}
\affiliation{Max-Planck-Institut f\"ur Quantenoptik, Hans-Kopfermann-Str. 1, D-85748 Garching, Germany}

\maketitle

\noindent
{\bf \small THIS PDF FILE INCLUDES:}\\
\indent Materials and Methods\\
\indent Supporting Text\\
\indent Figs.~S1-S7\\
\indent Table~S1-S4\\
\indent Reference~38-48\\ \\

In this supplement, we summarize the following additional information: In Sec.\,S.I we give a short description of the principle and design of the centrifuge decelerator. Sec.\,S.II summarizes the sources and purities of all compounds used in the experiment. In Sec.\,S.III we discuss the stability of the cryofuge at its maximal output intensity of cold molecules. In Sec.\,S.IV we provide details of the experimental conditions and show results for producing cold methanol and isopropanol beams from our cryofuge. Section\,S.V explains the procedure for tuning the molecular densities in the collision measurements, and the resulting systematic errors. The modeling and cancellation of these systematic errors are discussed in Sec.\,S.VI and Sec.\,S.VII. The systematic errors resulting from the QMS (Quadrupole Mass Spectrometer) detection due to space-charge effects and how they are eliminated are discussed in Sec.\,S.VIII. Section\,S.IX explains the scaling of the QMS sensitivity calibration for different species. Calculations of the loss probability for elastic collisions in the guide are discussed in Sec.\,S.X. Calculations of various collision cross-sections and loss rate coefficients are presented in Sec.\,S.XI. The effective length for collisions inside the centrifuge is elaborated in Sec.\,S.XII.

\section{Concept and design of the centrifuge decelerator}
\begin{figure}
	\centering
		\includegraphics[width=0.45\textwidth]{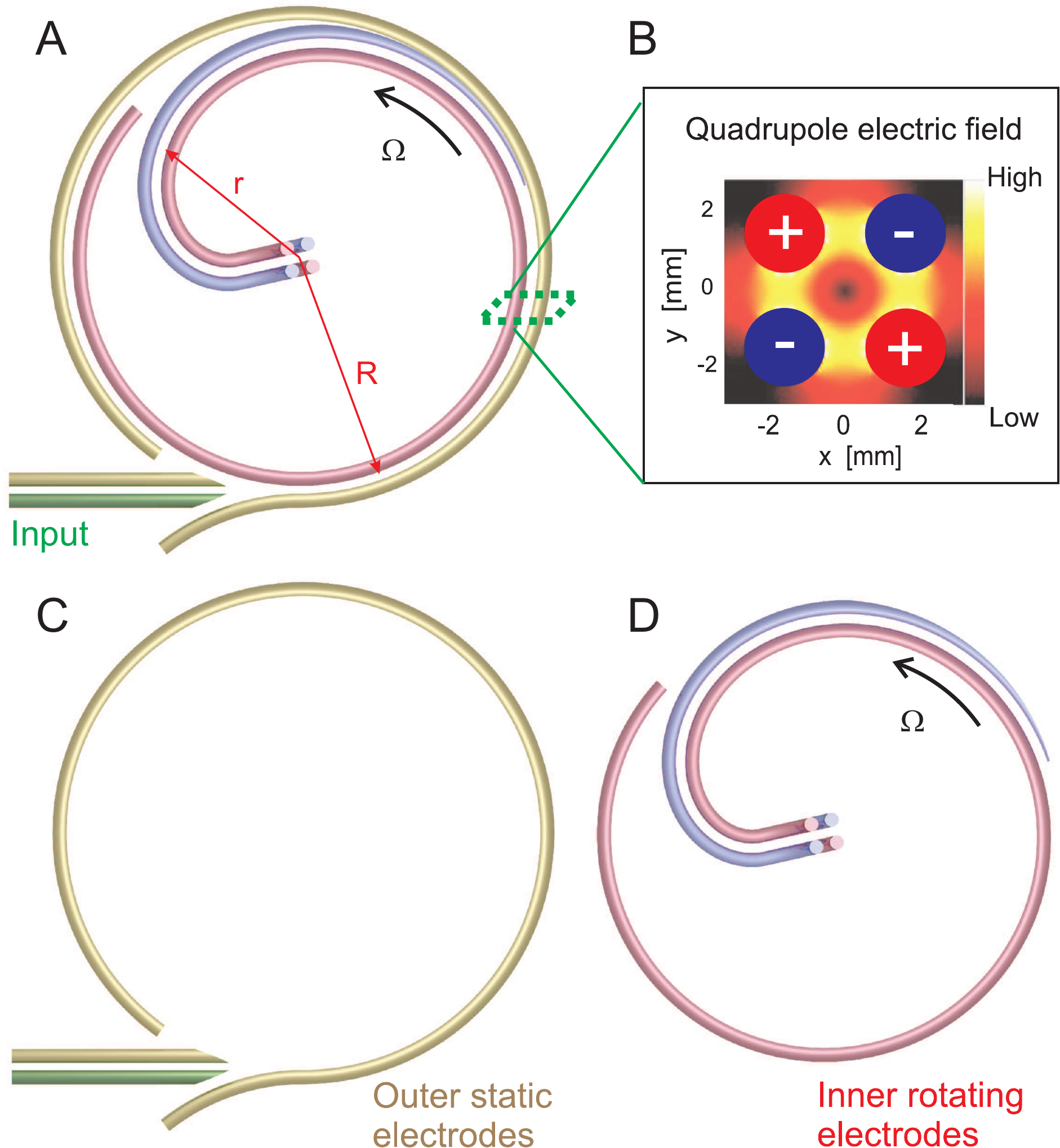}
	\caption{\textbf{Schematic of the centrifuge decelerator (top-view).} (\textbf{A}) Design assembly. (\textbf{B}) Transverse electric-field distribution in the quadrupole guide. (\textbf{C}) Static electrodes. (\textbf{D}) Rotating electrodes.}
	\label{fig:designcentrifuge}
\end{figure}

A basic idea for decelerating any particle beams is to make them climb up a potential hill at the expense of their kinetic energy. To produce a sufficient potential energy for decelerating continuous molecular beams at velocities up to $\sim200$\,m/s, we employ a non-inertial system, in particular harnessing the centrifugal force in a rotating frame (Fig.\,S1\,A). The centrifugal potential energy which is responsible for the deceleration in the rotating frame is given by $E_{centrifugal}=-\frac{1}{2}m(\vec{\Omega}\times\vec{r})^2$~\cite{Chervenkov2014}, where $m$ is the mass of the particle, $\vec{\Omega}$ is the angular velocity of the rotating frame, and $\vec{r}$ is the radial vector pointing from the center of the rotation to the position of the particle. A particle traveling from the periphery ($r=R=20$\,cm, the radius of the centrifuge) to the center of the rotating frame ($r = 0$\,cm) climbs up a potential hill of $\Delta E = m\Omega^2R^2/2$. This decrease in a particle's kinetic energy can be controlled by varying $\left|\vec{\Omega}\right|$. During the deceleration, the transverse divergence of the molecular beam is suppressed by quadrupole electrostatic guiding (Fig.\,S1\,B).

The design of the centrifuge enables a continuous deceleration of molecular beams~\cite{Chervenkov2014}. For this purpose, the decelerator consists of two parts: static electrodes around the periphery (Fig.\,S1\,C) and rotating electrodes in the inner region (Fig.\,S1\,D). The rotating guide has a spiral shape and consists of four parallel electrodes in a quadrupole configuration. The two inner rotating electrodes (red curve in Fig.\,S1\,D) extend in a circular shape parallel to the outer static electrodes. This ‘sweeping tail’ together with the two outer static electrodes form a storage ring of molecules around the periphery. This storage ring enables guiding of molecules around the periphery until they catch up with and enter the rotating spiral, irrespective of the moving position of the spiral's entrance, which ensures the deceleration of continuous beams. Further details on the centrifuge are presented in~\cite{Chervenkov2014}.

\section{Sources and purities of all compounds}
All compounds used in the experiment (including buffer-gas atoms) that are in gas phase at room temperature are from lecture bottles or gas cylinders. Methanol and Isopropanol are from spectroscopy grade solvents. The purities and commercial suppliers for all compounds are listed in Table\,S1.

\begin{table}[h]
	\centering
		\begin{tabular}{|c|c|c|}
			\hline
		 \ \ \textbf{Compound}\ \		&	\ \ \ \ \textbf{Purity}\ \ \ \ &	 \ \ \ \textbf{Commercial suppliers}\ \ \ 	\\ \hline
		 Helium	& $99.996\%$	& Westfalen \\ \hline
		 Neon 	& $99.995\%$	& Linde  \\ \hline
		 CH$_3$F 	& $99.5\%$	& Linde  \\ \hline
		 CF$_3$CCH 	& $\geq97\%$	& SynQuest Laboratories  \\ \hline
		 ND$_3$ 	& $99\%$	& Sigma-Aldrich, Spectra Gases  \\ \hline
		 Methanol 	& $99.8\%$	& Merck  \\ \hline
		 Isopropanol 	& $99.9\%$	& Merck  \\ \hline
		\end{tabular}
	\caption{
	\textbf{Purities of all compounds used in the experiment.}
	}
\end{table}
\section{Long-term stability of the cryofuge output}
To produce the flux of $(1.2\pm^{1.2}_{0.6})\times10^{10}$\,s$^{-1}$ and density of $(1.0\pm^{1.0}_{0.5})\times10^9$\,cm$^{-3}$ CH$_3$F molecules below $1$K$\times k_B$ kinetic energy in the laboratory frame, as quoted in the main text, the cryogenic buffer-gas cooling is operated with a $6.5$\,K cell, $0.5$\,sccm He inflow, and $0.2$\,sccm molecule inflow. The centrifuge runs at $30$\,Hz. The cryofuge operates stably for at least $6$\,hours, with $30$\% overall signal decrease due to ice formation  at the buffer-gas cell output. By reducing the molecule inflow from $0.2$\,sccm to $0.1$\,sccm, the slow flux obtained is reduced by $1/3$, and the stable operation time of the system increases to over $12$\,hours. Longer operation time can be achieved by implementing de-icing at the cell output with a pulsed heating current~\cite{Sommer2009}. In addition, the system can be reset by warming up the cryogenic part to above $100$\,K overnight.

\section{Cold methanol and isopropanol beams}
In this section, we provide the experimental conditions under which the intense beams of cold and slow methanol and isopropanol are produced, explain how we calculate the beam fluxes, and show the velocity distribution and TOF signal. For methanol, the buffer-gas cell is kept at $25$\,K, and the molecule-line inlet is heated up to $400$\,K to prevent freezing. The input fluxes of helium and methanol to the buffer-gas cell are stabilized to $0.2$\,sccm and $0.08$\,sccm, respectively. The centrifuge rotates at $37$\,Hz. For isopropanol, the buffer-gas cell temperature is $21$\,K, the molecule-line inlet temperature is heated to $300$\,K, the buffer-gas flux is set to $1.0$\,sccm and the centrifuge rotation frequency is set to $40$\,Hz.


The measured longitudinal velocity ($v_z$) distribution for methanol beams after the cryofuge is shown in Fig.\,S2\,A. The vertical axis in Fig.\,S2\,A represents the count rate per velocity interval. The $1\sigma$ errorbar in the figure is derived from signal short noise averaged over about $1.3$\,hours measurement time. The count rate is proportional to the molecular density at the detector, and the ratio between the two defines the calibration factor of the detector which is about $200$\,cm$^{-3}$\,per count/s for methanol in our case. The product of the area under the data points in Fig.\,S2\,A and the calibration factor gives the density at the detector. Multiplying this density by the averaged velocity and the beam spread (about $0.3$\,cm$^2$) which results from the divergence of molecules after they leave the guide, gives the beam flux, which is about $3\times10^8$\,s$^{-1}$ for the cold methanol. The same procedure is applied to obtain the beam flux of other cold species from the cryofuge. The internal-state purity of such a methanol beam is substantially higher than the one corresponding to a thermal distribution at $25$\,K (the temperature of the buffer-gas cell). This is due to the electrostatic filtering, as only molecules in states with sufficient Stark shifts can be guided.

We also show in Fig.\,S2\,B the TOF signal for the cold isopropanol beam. The vertical axis shows the count rate of the beam as a function of the arrival time at the detector. The maximal height of the signal minus the background gives the total signal for the isopropanol, which is about $900$\,cnt/s. The detector calibration factor for isopropanol is about $96$\,cm$^{-3}$\,per count/s. Thus following the same calculation mentioned above, we obtain the value for the isopropanol flux which is about $1\times10^8$\,s$^{-1}$.

\begin{figure}
	\centering
		\includegraphics[width=0.5\textwidth]{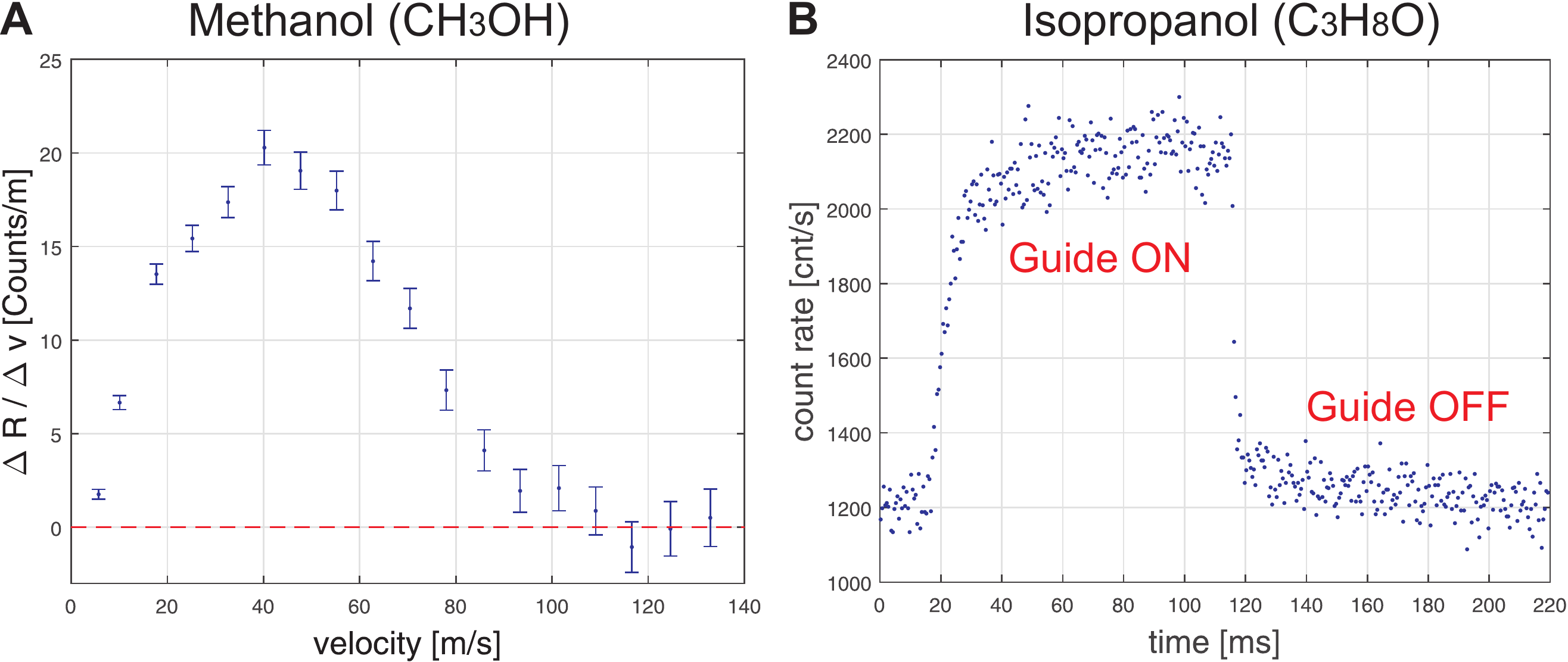}
	\caption{\textbf{Longitudinal velocity distribution for the cold methanol beam (A) and the TOF signal for the cold isopropanol beam (B).} The errorbars in (\textbf{A}) as well as in the figures and text through out the rest of the supplement represent $1\sigma$ statistical error of the measurements.}
	\label{fig:methanol}
\end{figure}

\section{Tuning the molecular density for the collision measurements}
One difficulty in studying the molecular density, $n$, dependent collision loss in the TOF guide (see Fig.\,1 of the main text), is to find methods for varying $n$ while maintaining all other parameters unchanged. Two approaches are applied in our measurement. First, we vary the voltage on the straight guide before the centrifuge decelerator, hence its transverse trap depth. In this way, ideally we can control the molecular density without changing its $v_z$-distribution, as $v_z$ and $v_{\perp}$ (the transverse velocity) are decoupled in the straight guide. Second, we regulate the molecular flux into the buffer-gas cell. During both operations, other parameters such as the centrifuge rotation speed, the buffer-gas cell temperature, and the buffer-gas input flux all stay unchanged. Notably, the voltage on the TOF guide is also kept constant, as otherwise it would simultaneously alter both $n$ and the trap depth in the collision region, as well as introduce changes to the molecular beam spread before the molecules arrive at the detector.

These two approaches, however, do not completely eliminate all side effects. The alteration of the transverse trap depth also modifies the measured $v_z$-distribution due to the coupling between the transverse and the longitudinal filtering in the bent guide before the straight segment, as explained in detail in Sec.\,S.VI. The tuning of the molecule inflow shifts the output velocity distribution due to the boosting effect~\cite{MotschBoosting} at the vicinity of the cell output aperture. This effect is small since the boosting is dominated by the density of the helium atoms, and it can be easily canceled out as explained in Sec.\,S.VII.

\section{The `collision-independent' background in the ratios of $v_z$-distributions}
\begin{figure}
	\centering
		\includegraphics[width=0.50\textwidth]{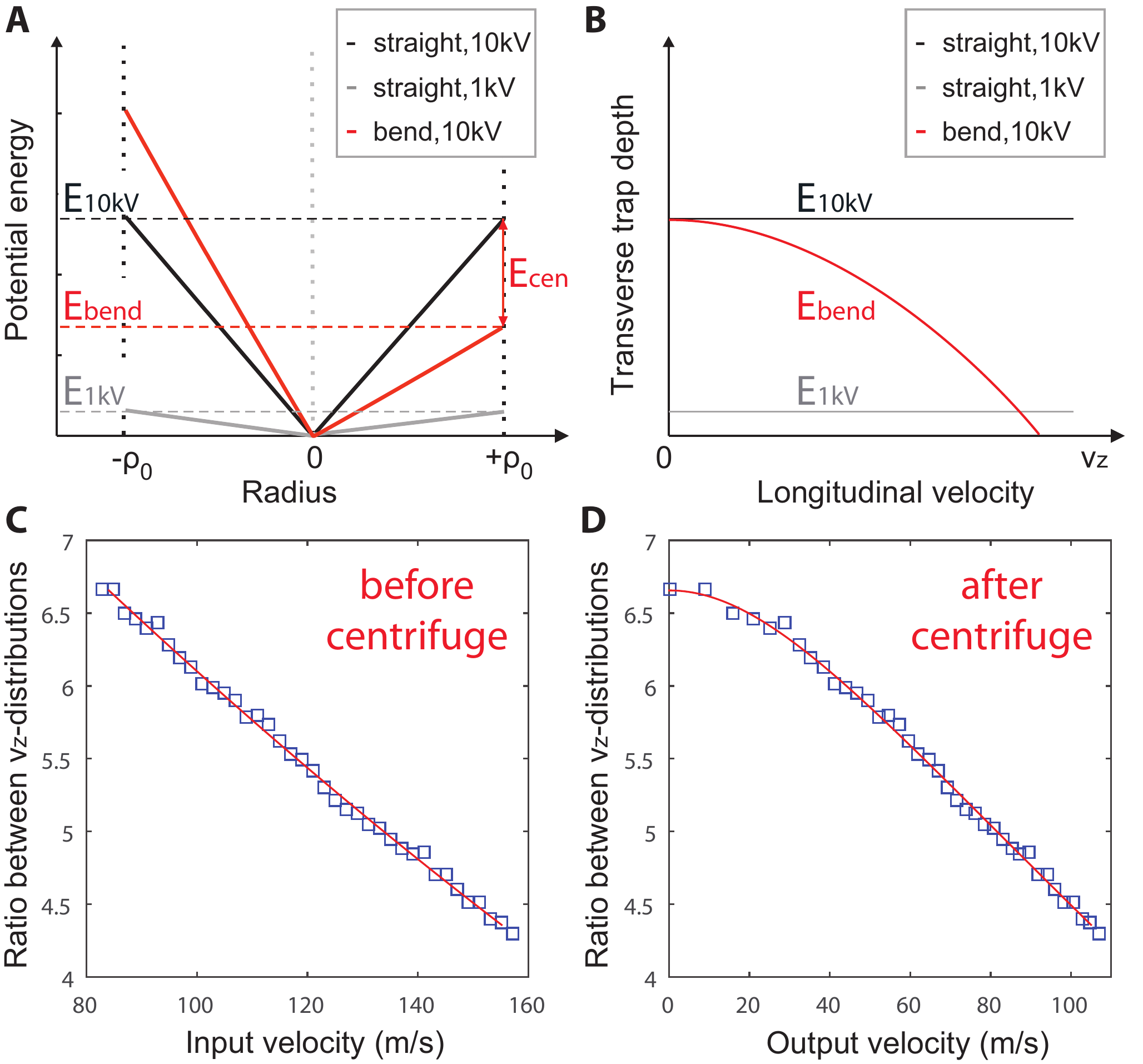}
	\caption{
	\textbf{Sketch of the $v_z$-dependent filtering on the transverse energy and the simulation results.} (\textbf{A}) Transverse potential energy of a molecule in a quadrupole guide assuming a linear Stark shift, at $10$\,kV (black), at $1$\,kV (gray), and at a bend with $10$\,kV (red) for a given longitudinal velocity. The dashed lines indicate the corresponding trap-depth. $0$ on the x-axis represents the guide center, and $\pm \rho_0$ represent the boundary of the guide. \textbf{B}) Transverse trap-depth as a function of $v_z$, for a straight guide (black and gray), and for a bent guide (red). (\textbf{C}) Ratio between the $v_z$ distributions at $10$\,kV and at $1$\,kV guiding voltage before the centrifuge, and (\textbf{D}) the same ratio after the centrifuge, which is operated at a rotation speed of $33.5$\,Hz. Blue squares show results from Monte Carlo trajectory simulations, and the red lines are smoothing curves. 
	}
	\label{fig:bendguide1}
\end{figure}

As mentioned in Sec.\,S.V, when we change the molecular density by varying the guiding voltage (e.g. toggling between $10$\,kV and $1$\,kV as carried out in our measurements) on the straight guide before the centrifuge, a collision-independent background is expected in the ratio of the obtained $v_z$-distributions. This is due to the coupling between the longitudinal and the transverse confinement of molecules in the bent quadrupole guide before this straight segment. For a straight quadrupole guide with inner radius $\rho_0$, the transverse potential energy vs. position for $10$\,kV and $1$\,kV are plotted in Fig.\,S3\,A, and are independent of $v_z$ (Fig.\,S3\,B). However, for a bend with a radius of curvature $R$ ($R\gg\rho_0$), the transverse trap depth ($E_{bend}$) is shifted by the centrifugal energy $E_{cen}=-mv_z^2\rho_0/R$. This effectively lowers the transverse trap depth of the bent guide by an amount equal to $E_{cen}$, hence $(E_{bend}-E_{bend}(v_z=0))\propto-v_z^2$ in Fig.\,S3\,B. Therefore, when the voltage on the straight segment after the bend is reduced from $10$\,kV to $1$\,kV, the loss in molecular signal must be $v_z$-dependent, as the bend has already effectively performed a $v_z$-dependent pre-filtering. 

To quantitatively predict this background which is independent of any collisions in the guide, we perform a complete Monte-Carlo trajectory simulation of the guiding from the output of the buffer-gas cell to the input of the centrifuge. The detailed description of this simulation method and its validation with a state-resolved measurement has been reported previously~\cite{Wu2016}. The simulated point-by-point ratio between the velocity distributions before entering the centrifuge is plotted in Fig.\,S3\,C. The transformation from the input $v_{in,z}$ to the output $v_{out,z}$ of the centrifuge is simply given by $v_{out,z}=\sqrt{v_{in,z}^2-2v_pv_{in,z}}$, where $v_p=2\pi\Omega R'$ is the peripheral velocity of the centrifuge, the rotation speed is $\Omega=33.5$\,Hz for the measurement shown in Fig.\,4\,A in the main text, and the radius of our centrifuge is $R'=20$\,cm. The transformed ratio of output $v_z$-distributions is plotted in Fig.\,S3\,D. The experimentally obtained ratio of $v_z$-distributions from the collision measurements are compared to this simulated ratio, as it represents the model where no collisions occur. We perform separate simulations for CH$_3$F and ND$_3$, since they are investigated under different experimental conditions.

\section{Cancellation of the boosting effect in the ratios of $v_z$-distributions}
The second way of varying the molecular density in the TOF-guide is by directly changing the molecule inflow to the buffer-gas cell. This however also introduces a small shift to the velocity distribution of the molecules leaving the buffer-gas cell, due to the boosting effect~\cite{MotschBoosting}, which is difficult to simulate. Thus one cannot directly take the ratio of $v_z$-distributions measured at different molecule inflow rates. To cancel this shift out, at each given inflow we vary the voltage on the straight guide before the centrifuge (the first approach in Sec.\,S.V), and then take the point-by-point ratio of the $v_z$-distributions measured at these two voltages. The boosting effect cancels out in the ratio, as it is the same for both measurements. By changing to a different molecule inflow rate while keeping the same voltage switching procedure, we can tune the value of $\Delta n$, as shown in Fig.\,4\,A.

\section{Linear response of the detector} 
In addition to canceling or compensating the aforementioned effects resulting from the density control, we have also taken full care of the systematics of the detector. We measure molecules with a cross-beam quadrupole mass spectrometer (QMS 410, Pfeiffer) which generates electron beams in the detection volume, ionizes the incoming molecules via electron impact, and then collects and counts the ions. For such a system, it is known from literature~\cite{Beijerinck1974} that systematics in the temporal response could arise from the space-charge effect. When the electron cloud in the detection volume is sufficiently dense, the resultant electrostatic interaction disturbs the extraction of the produced ions, and subsequently causes a delay in the arrival time at the ion counter~\cite{BRAUN1993}. As a result, the measured $v_z$-distribution would appear slower than the actual distribution. To avoid such artifacts, we have to reduce the electron emission current until a linear response from the TOF measurement is obtained. In this way, we make sure that the measured $v_z$-profile becomes independent of the emission current (Fig.\,S4\,A). Consequently, all the measurements on $v_z$-distributions and collision effects reported in the main text were performed at $60$ to $120\,\mu$A emission current, instead of the typical range of $400$ to $600\,\mu$A recommended for a general application. The drawback of reducing the emission current is of course the decrease in the detection efficiency. However, the signals we obtain are still sufficient to reveal the deceleration and collisions thanks to the large molecular intensity produced by the cryofuge.
\begin{figure}
	\centering
		\includegraphics[width=0.50\textwidth]{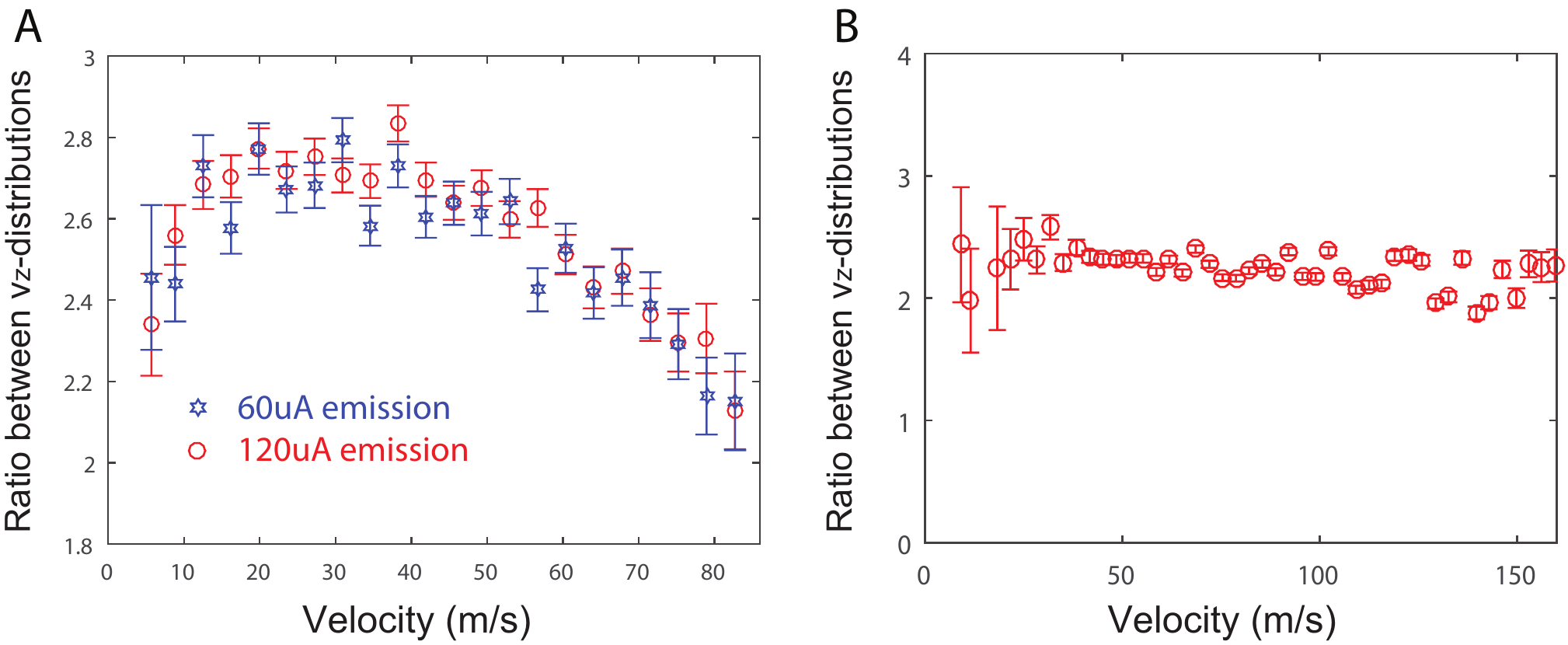}
	\caption{
	\textbf{Verification of the linear response of the detector.} (\textbf{A}) Ratios of $v_z$-distributions in the collision measurements taken at $60$ and $120\,\mu$A emission current. All other parameters are identical, namely $0.5$\,sccm He, $0.25$\,sccm CH$_3$F, $33.5$\,Hz centrifuge rotation, and the voltage on the straight guide before the centrifuge is toggled between $10$ and $1$\,kV. The data are averaged over $2.5$ and $1$\,hour for the $60$ and $120\,\mu$A measurements, respectively. (\textbf{B}) Ratio between $v_z$-distributions taken at different distances away from the end of the TOF guide, $24$\,mm over $36$\,mm. This test measurement was performed on a different setup~\cite{Sommer2010} where molecules from a $120$\,K effusive nozzle are directly guided to the detector.
	}
	\label{fig:linearity}
\end{figure}

Apart from the dependence on the emission current, we have also verified the linear response of the QMS on the ion density. When the ion density produced in the detection volume is sufficiently high, one could expect a similar distortion of the ion trajectory, and hence their extraction and arrival at the ion counter. To rule out this nonlinearity, we varied the distance from the end of the TOF guide to the QMS detection volume while keeping the incoming molecular intensity the same. Due to the beam divergence, the molecular density reaching the QMS scales inversely to the distance squared. This effectively changes the ion density present in the detection volume. According to the results from test measurements (Fig.\,S4\,B), our detector still exhibits a linear response within the relevant density range.    

\begin{table}
	\centering
		\begin{tabular}{|c|c|c|c|c|}
			\hline
		 \ \ \textbf{Species}\ \		&	\ \ \ \ \textbf{N$_2$}\ \ \ \ &	 \ \ \ \textbf{ND$_3$}\ \ \ 	&  \ \ \textbf{CH$_3$F}\ \ 	&  \textbf{CH$_3$OH}\ \\ \hline
		 ionization cross section [\AA$^2$]	& $2.51$	& $3.01$	& $3.41$	&$4.69$  \\ \hline
		 static polarizability [\AA$^3$]		&  1.71					& 2.10								  & 2.54	& 3.21\\ \hline
		\end{tabular}
	\caption{
	\textbf{List of electron impact ionization cross sections and static polarizabilities of various species, taken from literature.} The cited ionization cross sections are for electron impact energy of $70$\,eV, and the corresponding references are N$_2$\cite{Hwang1996}, ND$_3$\cite{Rao1992}, CH$_3$F\cite{Torres2001}, CH$_3$OH\cite{Srivastava1996}. The values for the static polarizability are obtained from~\cite{nistdatabase}.
	}
\end{table}
\section{Dependence of QMS sensitivity on the molecular species}
The calibration of the QMS sensitivity has been performed only for CH$_3$F, and rescaled for the other species based on the literature values of their electron impact ionization cross section (Table\,S2). Consequently, the systematics in the density calibration should cancel out in the ratio of the measured collision rates $k_{loss}^{ND_3}/k_{loss}^{CH_3F}$ as discussed in the main text. In addition, the electron impact ionization cross-section is found to be roughly proportional to the molecular polarizability~\cite{Lampe1957}, in case the literature value for a specific molecule species is not available.

\section{Loss probability from elastic collisions in a quadrupole guide}

In this section, we summarize the method for computing the probability of losses induced by elastic collisions. More specifically, our goal is to compute, upon experiencing an elastic collision, the probability for a molecule to get lost from the quadrupole guide (the TOF-segment) for the given trap depth, and for any scattering angle $\theta$ and relative longitudinal velocity $v_{z}^{rel}$. The dependence on the transverse velocity and position distributions in the guide are averaged over the whole ensemble in the calculation. The trap depth in our case is about $0.9$K$\times k_B$ for CH$_3$F and $1.2$K$\times k_B$ for ND$_3$.

\begin{figure}
	\centering
		\includegraphics[width=0.5\textwidth]{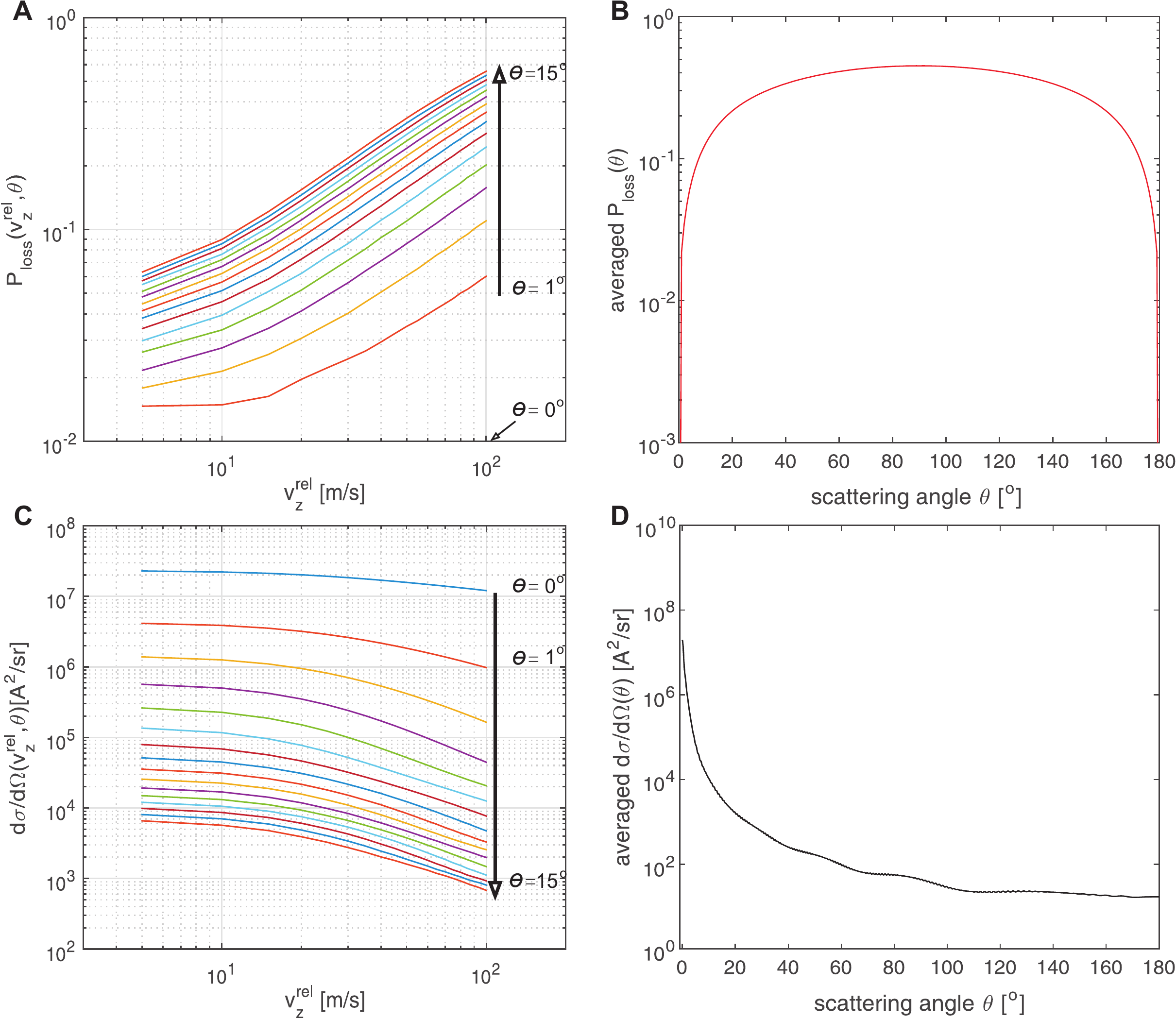}
	\caption{
	\textbf{Angular and velocity dependence of the loss probability and the differential cross section of elastic scattering between CH$_3$F molecules in the quadrupole guide}. (\textbf{A}) Simulated loss probability $P_{loss}(v_z^{rel},\theta)$ as a function of $v_z^{rel}$ for selected scattering angles (from $0^o$ to $15^o$, and one curve per degree). (\textbf{B}) Loss probability $P_{loss}(\theta)$ as a function of the scattering angle, after averaging over the measured $v_z$-distribution. (\textbf{C}) Calculated differential cross section $\frac{d\sigma}{d\Omega}(v_z^{rel},\theta)$ of elastic collisions as a function of $v_z^{rel}$ for selected scattering angles (from $0^o$ to $15^o$, and one curve per degree), from the eikonal approximation. (\textbf{D}) Differential cross-section $\frac{d\sigma}{d\Omega}(\theta)$ as a function of the scattering angle, after averaging over the measured $v_z$-distribution.
	}
	\label{fig:LossProbability}
\end{figure}

First of all, we need to work out the energy and spatial distribution of molecules in the guide. The transverse energy and spatial distribution can be obtained from Monte-Carlo trajectory simulations~\cite{Wu2016}. For simplicity, assuming a perfect linear electric field distribution in the quadrupole guide, and a perfect linear Stark-shift experienced by molecules in the guide, the transverse energy distribution can be worked out from a phase-space volume consideration, and is proportional to $v_{\perp}^5$, where $v_{\perp}$ is the transverse velocity of a molecule. The transverse spatial distribution of a single molecule can be approximated with $R(\rho)\propto \rho(\rho_0-\rho)$, where $\rho$ is the radial coordinate and $\rho_0\approx1$\,mm is the inner radius of the guide. The transverse spatial distribution for a collision between two molecules is then $R_{col}(\rho)\propto R^2(\rho)/\rho=\rho(\rho_0-\rho)^2$, where $\rho$ in the denominator is to divide out the probability for the colliding partners to have different azimuthal positions.

Next, we use random number generation to produce an initial ensemble with the position and energy distributions obtained from the first step. The calculation of the energy transfer upon one elastic collision is then performed in the center-of-mass frame, for a given pair of relative longitudinal velocity $v_{z}^{rel}$ and scattering angle $\theta$, where the relative transverse velocity $v_{\perp}^{rel}$ is taken from the entire ensemble and the azimuthal angle of the collision covers a flat distribution from $0$ to $2\pi$. The final velocity after the collision is then transformed back to the laboratory frame.

As a last step, we compare the final total transverse energy (kinetic + potential) in the laboratory frame with the trap depth of the guide. A molecule is lost if its transverse energy exceeds the trap depth. The percentage of lost molecules in the whole ensemble gives the loss probability $P_{loss}(v_{z}^{rel},\theta)$ for the given $v_{z}^{rel}$ and $\theta$, but averaged over the transverse spatial and energy distributions. The calculation for CH$_3$F is plotted in Fig.\,S5\,A. Note that in principle both molecules have the probability to get lost during one collision event, and our model only includes the loss probability of one of the two colliding partners, to avoid double counting. Two features are apparent in Fig.\,S5\,A. First, the smaller the scattering angle, the lower the loss probability for a given $v_{z}^{rel}$, and $P_{loss}(v_{z}^{rel},\theta=0^o)=0$ as there is no energy transfer from the longitudinal to the transverse component of the molecule in this limit. Second, the smaller the $v_{z}^{rel}$, the lower the loss probability for a given $\theta$, as less energy can be transferred to the transverse direction in a single collision.

By integrating $v_{z}^{rel}$ over our measured longitudinal velocity distribution, we obtain the averaged loss probability $P_{loss}(\theta)$ as a function of the scattering angle (Fig.\,S5\,B). The corresponding averaged collision energy is $0.8$K$\times k_B$ for CH$_3$F in our measurement (Table\,S3). The plot in Fig.\,S5\,B is consistent with the features observed in Fig.\,S5\,A. Firstly, at $\theta<10^o$, the loss probability quickly drops to zero. Secondly, the maximal loss probability is only $45$\% at $\theta$ around $90^o$.   

\begin{table}
	\centering
		\begin{tabular}{|c|c|c|}
			\hline

		 \textbf{Collision energy}						   &	\ \ \textbf{CH$_3$F} \ \  						&\ \ \ \textbf{ND$_3$} \ \ \ \			\\ 					\hline
		 transverse $\bar{v}_{\perp}^{rel}$ [m/s]&  $17.3$												&				$26.4$												\\					\hline
		 longitudinal $\bar{v}_{z}^{rel}$ [m/s]  &  $22.3$												&				$33.2$												\\					\hline
		 								 $E_{col}/k_B$ [K]			 &  $0.8$											&				$1.1$     										\\					\hline

		\end{tabular}
	\caption{
	\textbf{List of averaged velocities and collision energies for CH$_3$F-CH$_3$F and ND$_3$-ND$_3$ collisions in our measurement.} $\bar{v}_{\perp}^{rel}$ is the relative transverse velocity averaged over the known distribution. $\bar{v}_{z}^{rel}$ is the relative longitudinal velocity averaged over the measured $v_z$-distribution. The collision energy is $E_{col}=\mu\bar{v}_{rel}^2/2$, where $\mu$ is the reduced mass, and the relative velocity $\bar{v}_{rel}=\sqrt{(\bar{v}_{z}^{rel})^2+(\bar{v}_{\perp}^{rel})^2}$.}
\end{table}

\section{Calculation and fitting of the scattering cross-section and rate coefficient}
The collision loss cross section and rate coefficient have contributions from both elastic and inelastic channels. The loss probability for elastic collisions $P_{loss}(v_{z}^{rel},\theta)$ is calculated in Sec.\,S.X. The differential cross section $\frac{d\sigma}{d\Omega}(v_{z}^{rel},\theta)$ is computed from the semiclassical eikonal approximation~\cite{SakuraiQM}, taking into account the isotropic part of the dipolar interaction $V_{dd}(r)=-\left\langle d\right\rangle^2/4\pi\epsilon_0r^3$. The total relative velocity $v_{rel}$, which is the directly relevant variable for computing $\frac{d\sigma}{d\Omega}$, is calculated from $v_{z}^{rel}$ and $\bar{v}_{\perp}^{rel}$ averaged over the whole transverse distributions. The results are plotted in Fig.\,S5\,C for $0^o$ to $15^o$ scattering angle. Averaging over all $v_{z}^{rel}$ for the measured velocity distribution gives $\frac{d\sigma}{d\Omega}(\theta)$, and is plotted as a function of $\theta$ in Fig.\,S5\,D. The plot shows clearly that $\frac{d\sigma}{d\Omega}(\theta)$ decreases quickly as $\theta$ increases. This is because in the semiclassical regime the elastic scattering concentrates in the forward direction. As the loss probability at smaller scattering angles is low (Fig.\,S5\,B), most elastic collisions do not lead to losses.

The elastic loss cross section as a function of the relative velocity is then computed from the integral $\sigma^{el}_{loss}(v_{z}^{rel})=\int d\phi sin\theta d\theta P_{loss}(v_{z}^{rel},\theta)\frac{d\sigma}{d\Omega}(v_{z}^{rel},\theta)$. The corresponding loss rate coefficient is simply $k^{el}_{loss}(v_{z}^{rel})=\sigma^{el}_{loss}(v_{z}^{rel})v_{rel}(v_{z}^{rel})$ (blue curve in Fig.\,S6). The total elastic rate coefficient is obtained from the integration without including the loss probability, $k^{el}_{total}=\sigma^{el}_{total}v_{rel}=v_{rel}\int d\phi sin\theta d\theta \frac{d\sigma}{d\Omega}$. As shown in Fig.\,S6 (red curve), $k^{el}_{total}$ is independent of the collision energy in the semiclassical limit. The inelastic loss cross section and rate coefficient are estimated from the Langevin capture model, which is summarized in~\cite{Bell2009}. The Langevin rate for dipolar scattering is $\propto v_{rel}^{-1/3}$. The results for $k^{el}_{loss}$, $k^{el}_{total}$, $k^{in}$, and the theoretical total loss rate coefficient $k_{th}=k^{el}_{loss}+k^{in}$ for our CH$_3$F beam, are plotted as functions of $v_{z}^{rel}$ in Fig.\,S6. 

With the measured $v_z$-distribution of the molecules in the TOF-guide, we can transform $k_{th}(v_{z}^{rel})$ into $k_{th}(v_z)$, and fit it to the experimental data using the $exp(-\alpha k_{th}\Delta nL/v_z)$-model (Fig.\,4), where $\Delta n$ is the measured density difference, and $L=L_{TOF}+L_{eff}$ is the guide length with the main contribution $L_{TOF}=46$\,cm for the TOF-guide and an additional effective length inside the centrifuge $L_{eff}$ (see Sec.\,S.XII). The fitting parameter $\alpha$ accounts for the deviation between theory and experiment, which is about $1.4$ (see main text). The measured rate constant $k_{loss}^{CH_3F}$ is also plotted at the averaged $v_z^{rel}$ in Fig.\,S6 for comparison. In addition, we also list in Table\,S4 the various collision cross sections obtained from our measurement and from theory, for both CH$_3$F and ND$_3$.

\begin{figure}
	\centering
		\includegraphics[width=0.5\textwidth]{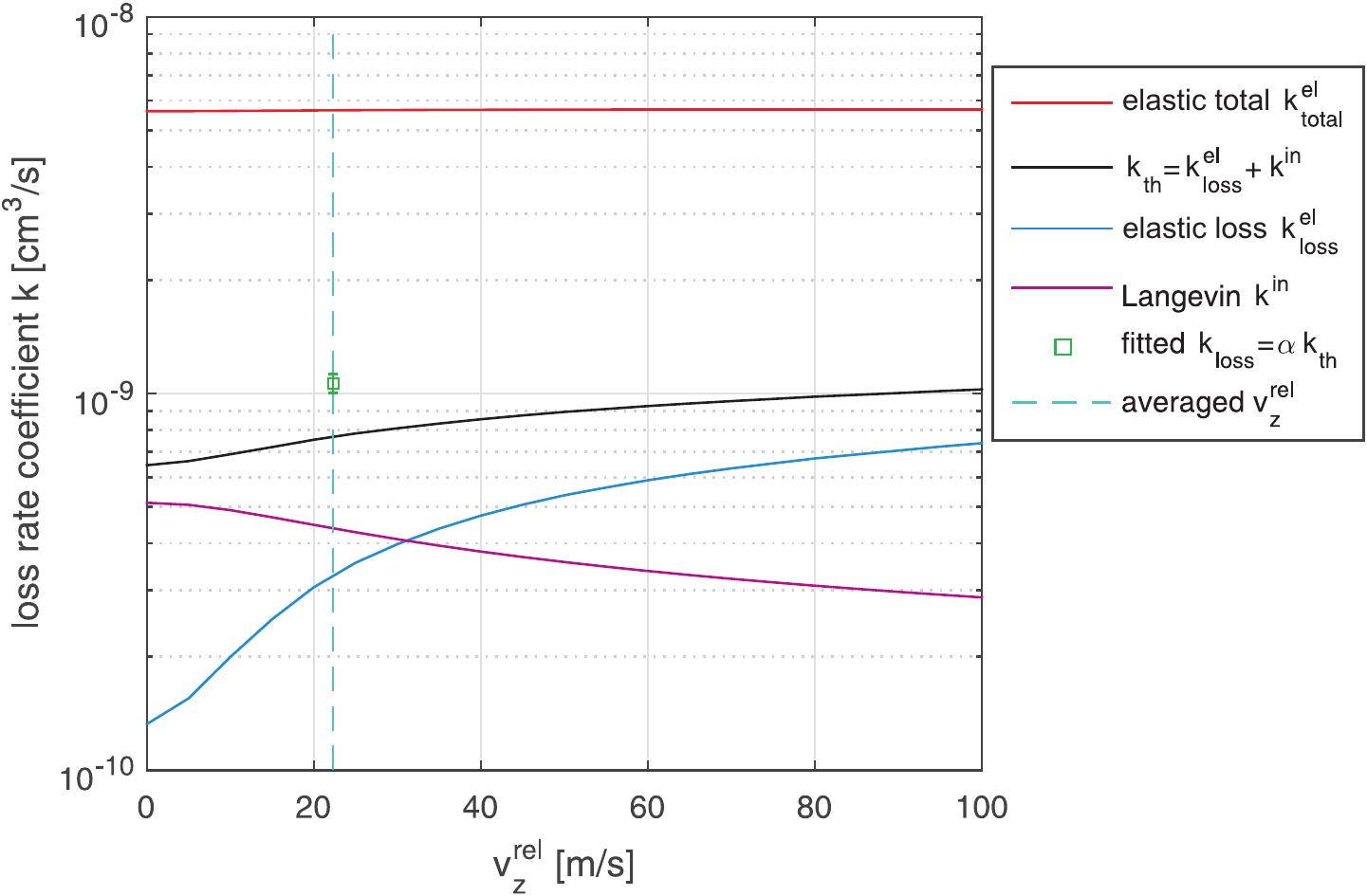}
	\caption{
	\textbf{Calculated collision rate coefficients and the experimentally fitted loss rate coefficient for the decelerated CH$_3$F beam}. 
	}
	\label{fig:Ratecoefficient}
\end{figure}

\begin{table}
	\centering
		\begin{tabular}{|c|c|c|}
			\hline
		 \ \ \textbf{Cross sections} \ \		&	\ \ \ \textbf{CH$_3$F} at \ \ \ &\ \ \ \textbf{ND$_3$} at \ \ \ 	\\ 
		 		($\times10^{-12}$\,cm$^2$)			&			$E_{col}/k_B=0.8$\,K		&		$E_{col}/k_B=1.1$\,K     \\	\hline
		 measured loss $\sigma_{loss}$ 			&  $0.38\pm0.03$					    &      $0.49\pm0.09$										\\ \hline
		 theoretical loss $\sigma_{th}$			&  $0.27$											& $0.31$								   							\\ \hline
		 Langevin $\sigma^{in}$							&  $0.15$											& $0.16$								   							\\ \hline
		 elastic loss $\sigma_{loss}^{el}$	&  $0.12$											& $0.14$								   							\\ \hline
		 elastic total $\sigma_{total}^{el}$&  $2.0$											& $2.5$								   								\\ \hline

		\end{tabular}
	\caption{
	\textbf{The list of various cross sections for CH$_3$F-CH$_3$F and ND$_3$-ND$_3$ collisions at the corresponding collision energies.} The measured loss cross section $\sigma_{loss}$ is given in the first row. The corresponding error bars are only statistical from the fitting of the data. It agrees with the theory $\sigma_{th}$ within $40$\%. From the second to the last row, we show theoretically calculated cross sections, including the theoretical loss cross section $\sigma_{th}=\sigma^{in}+\sigma_{loss}^{el}$, the inelastic cross section from Langevin model $\sigma^{in}$, the elastic loss cross section for our specific system $\sigma_{loss}^{el}$, and the total elastic cross section $\sigma_{total}^{el}$.}
\end{table}

\section{Effective collision length inside the centrifuge}
Collisions between molecules start to build up already inside the centrifuge before the molecules are fully decelerated. Neglecting this effect (i.e. setting the effective length inside the centrifuge $L_{eff}=0$) would introduce a systematic error in the length of the collision region $L$, and consequently would result in a greater deviation between the measured loss rate and the theoretical values (i.e. a value of $\alpha$ greater than $1.4$). By including $L_{eff}$ in the collision model, we obtain the single-particle loss factor $exp(-\alpha k_{th}n\left(L_{TOF}+L_{eff}\right)/v_z)=exp(-\alpha k_{th}n\left(L_{TOF}/v_z+T_{eff}\right))$, where the effective collision time in the centrifuge $T_{eff}=\tau\frac{n'}{n}\frac{k'_{th}}{k_{th}}$. Here, $\tau$ is the actual transient time in the rotating guide, $n'$ and $k'_{th}$ are the molecule density and collision loss rate inside the centrifuge, respectively, and $n$ and $k_{th}$ are the corresponding quantities inside TOF-guide.

The details of this correction term are explained in the following. $\tau$ can be calculated numerically since we know the exact shape of the spiral trajectory and the input longitudinal velocity $v_{in}$-distribution. From conservation of energy in the rotating frame, the intermediate velocity distribution at each step along the spiral can be obtained (Fig.\,S7\,A), and integrating throughout the trajectory gives $\tau$. Clearly, $\tau$ is a function of the output velocity $v_{exit}$, since the slower the molecules, the longer it takes for them to travel through the centrifuge. The variation of molecule density $n'$ inside the centrifuge comes from two main contributions, the conservation of flux in the rotating frame during deceleration and the filtering at the output of the centrifuge. The former results in an increasing density as molecules travel towards the center of the spiral while slowing down, assuming the electric guiding has unit efficiency. The latter is responsible for the major filtering loss because of the following two reasons. First, the bend at the centrifuge output has the sharpest radius of curvature ($5$\,cm), which has a stronger effect for fast molecules. Second, the sharp bend is followed by a $1.2$\,mm gap between the the rotating exit guide and the static TOF guide, which provides a loss channel for (very) slow molecules. Trajectory simulations of the output bend and gap combined (Fig.\,S7\,B) predict an efficiency of $66$\% for the final velocity distribution at the center of rotation. Finally, the loss rate $k'_{th}$ inside the centrifuge can be worked out as well, since the averaged relative longitudinal velocity ($\bar{v}_{z}^{rel}$) of the beam can be calculated from the intermediate velocity distributions, and the transverse velocity distribution can be calculated from the guiding potential of the electrodes. The final result of $T_{eff}$ versus the output velocity $v_{exit}$ is plotted in Fig.\,S7\,C, and the corresponding effective guide length inside the centrifuge $L_{eff}$ is plotted in Fig.\,S7\,D. For molecules exiting the centrifuge at about $10$\,m/s, which are responsible for the majority of the collision signal, the $L_{eff}$ is about $27$\,cm.

The main simplification made in this model is that we have neglected the filtering of molecules in the spiral guide. This effect should however be small, since molecules are pre-filtered before entering the spiral guide, and the spiral trajectory is designed to collect all longitudinal velocities after the pre-filtering. In addition, mixing between longitudinal and transverse velocities in the spiral guide can cause a small amount of transverse heating, hence lead to a small amount of loss of molecules from the guide. Neglecting the two effects introduce a slight underestimation of molecule density inside the centrifuge in the model.

\begin{figure}
	\centering
		\includegraphics[width=0.5\textwidth]{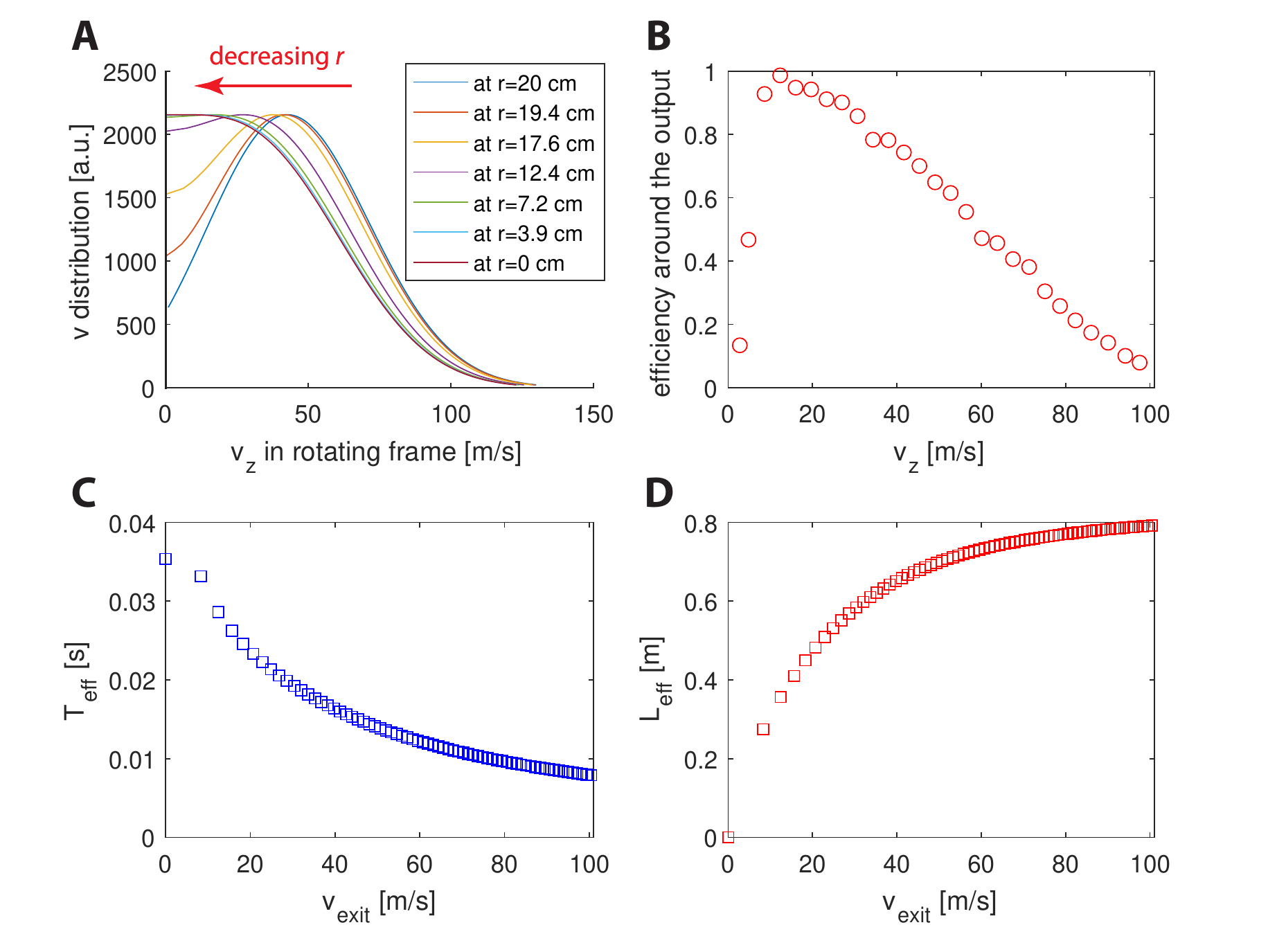}
	\caption{
	\textbf{Modeling the collision effects for CH$_3$F inside the centrifuge.} (\textbf{A}) Variation of $v_z$-distribution inside the centrifuge. The arrow indicates the trend of decreasing $r$, which is the distance to the center of rotation. (\textbf{B}) Guiding efficiency at the centrifuge output from trajectory simulations. The effects of both the $5$\,cm radius bend and the gap between centrifuge and the TOF-guide are included. (\textbf{C}) The effective collision time $T_{eff}$inside the centrifuge as a function of the output velocity, and (\textbf{D}) the corresponding effective guide length $L_{eff}$. 
	}
	\label{fig:insidecentrifuge}
\end{figure}


\bibliographystyle{unsrt}